\documentclass[preprint2]{aastex}

\usepackage{subfigure}
\usepackage{float}
\usepackage{color}
\usepackage[
    breaklinks=true,
    colorlinks=true,
    citecolor=blue,
    linkcolor=blue,
]{hyperref}
\usepackage{nameref}

\slugcomment{Version: \today}

\def\GIXa      {G029.86$-$00.04} 
\def\Jfour     {J1834$-$0301}
\def\JIII      {J1833$-$0323}
\def\Jseven    {J1857$-$0048}
\def\JVII      {J1827$-$0405}

\def\GIXb      {G029.95$-$00.01} 

\def\GIa       {G031.28$+$00.06} 
\def\Jthree    {J1853$-$0048} 

\def\GIb       {G031.58$+$00.07} 
\def\JIV       {J1904+0110}

\def\uas       {$\mu$as}
\def\deg       {$^\circ$}
\def\decdeg    {\ifmmode{{\rlap.}^{\circ}} \else ${\rlap.}^{\circ}$\fi}
\def\decs      {\ifmmode{{\rlap.}^{\rm s}} \else ${\rlap.}^{\rm s}$\fi}
\def\decas     {\ifmmode{{\rlap.}{''}}\else{${\rlap.}{''}$}\fi}

\def\kms       {km~s$^{-1}$}
\def\masy      {mas~yr$^{-1}$}
\def\jybeam    {Jy~beam$^{-1}$}

\def\meth      {CH$_3$OH}
\def\hho       {H$_2$O}
\def\VLSR      {$V_{\rm LSR}$}

\def\Lsun  {$L_\odot$}

\def\Msun  {$M_\odot$}

\def\mux   {\ifmmode {\mu_x}\else {$\mu_x$}\fi}
\def\muy   {\ifmmode {\mu_y}\else {$\mu_y$}\fi}
\def\mura  {\ifmmode {\mu_{\alpha}}\else {$\mu_{\alpha}$}\fi}
\def\mude  {\ifmmode {\mu_{\delta}}\else {$\mu_{\delta}$}\fi}

\def\dx    {$\Delta x$}
\def\dy    {$\Delta y$}

\def\Ts    {\ifmmode{\Theta_s}\else{$\Theta_s$}\fi}
\def\Tdot  {\ifmmode{d\Theta\over dR}\else{$d\Theta\over dR$}\fi}
\def\Rs    {\ifmmode{R_s}\else{$R_s$}\fi}
\def\To    {\ifmmode{\Theta_0}\else{$\Theta_0$}\fi}
\def\Ro    {\ifmmode{R_0}\else{$R_0$}\fi}

\def\Vo    {\ifmmode {V^{Std}_\odot}\else {$V^{Std}_\odot$}\fi}
\def\Uo    {\ifmmode {U^{Std}_\odot}\else {$U^{Std}_\odot$}\fi}
\def\Wo    {\ifmmode {W^{Std}_\odot}\else {$W^{Std}_\odot$}\fi}
\def\VH    {\ifmmode {V^H_\odot}\else {$V^H_\odot$}\fi}
\def\UH    {\ifmmode {U^H_\odot}\else {$U^H_\odot$}\fi}
\def\WH    {\ifmmode {W^H_\odot}\else {$W^H_\odot$}\fi}
\def\V     {\ifmmode {V_\odot}\else {$V_\odot$}\fi}
\def\U     {\ifmmode {U_\odot}\else {$U_\odot$}\fi}
\def\W     {\ifmmode {W_\odot}\else {$W_\odot$}\fi}
\def\VGC   {\ifmmode {V_\odot^{GC}}\else {$V_\odot^{GC}$}\fi}
\def\UGC   {\ifmmode {U_\odot^{GC}}\else {$U_\odot^{GC}$}\fi}
\def\WGC   {\ifmmode {W_\odot^{GC}}\else {$W_\odot^{GC}$}\fi}

\def\Vs    {\ifmmode {V_s}\else {$V_s$}\fi}
\def\Us    {\ifmmode {U_s}\else {$U_s$}\fi}
\def\Ws    {\ifmmode {W_s}\else {$W_s$}\fi}

\def\Vsbar {\ifmmode {\overline{V_s}}\else {$\overline{V_s}$}\fi}
\def\Usbar {\ifmmode {\overline{U_s}}\else {$\overline{U_s}$}\fi}
\def\Wsbar {\ifmmode {\overline{W_s}}\else {$\overline{W_s}$}\fi}

\newcommand{\HI}{\mbox{H\,\textsc{i}}}
\newcommand{\HII}{\mbox{H\,\textsc{ii}}}
\def\coIII {$^{13}$CO}
\def\coII  {$^{12}$CO}

\usepackage{ulem}

\begin{document}

\title{The Parallax of W43: a Massive Star Forming Complex near the
Galactic Bar}

\author{
 B. Zhang\altaffilmark{1,2},
 L. Moscadelli\altaffilmark{3},
 M. Sato\altaffilmark{1},
 M. J. Reid\altaffilmark{4},
 K. M. Menten\altaffilmark{1},
 X. W. Zheng\altaffilmark{5},
 A. Brunthaler\altaffilmark{1},
 T. M. Dame\altaffilmark{4},
 Y. Xu\altaffilmark{6} \&
 K. Immer\altaffilmark{1}
}

\altaffiltext{1}{Max-Plank-Institut f\"ur Radioastronomie, Auf dem H\"ugel
69, 53121 Bonn, Germany}

\altaffiltext{2}{Shanghai Astronomical Observatory, Chinese Academy of
Sciences, Shanghai 200030, China}

\altaffiltext{3}{INAF, Osservatorio Astrofisico di Arcetri, Largo E.
Fermi 5, 50125 Firenze, Italy}

\altaffiltext{4}{Harvard-Smithsonian Center for Astrophysics, 60
 Garden Street, Cambridge, MA 02138, USA}

\altaffiltext{5}{Department of Astronomy, Nanjing University, Nanjing
    210093, China}

\altaffiltext{6}{Purple Mountain Observatory, Chinese Academy of
Sciences, Nanjing 210008, China}

\begin{abstract}

We report trigonometric parallax measurements of masers in the massive
star forming complex W43 from VLBA observations as part of the BeSSeL
Survey. Based on measurements of three 12 GHz methanol maser sources
(\GIXa, \GIXb, and \GIa) and one 22 GHz water maser source (\GIb) toward
W43, we derived a distance of $5.49^{+0.39}_{-0.34}$~kpc to W43. By
associating the masers with CO molecular clouds, and associating the
clouds kinematically with CO longitude-velocity spiral features, we
assign W43 to the Scutum spiral arm, close to the near end of the
Galactic bar.  The peculiar motion of W43 is about 20 \kms\ toward the
Galactic Center and is very likely induced by the gravitational
attraction of the bar.

\end{abstract}

\keywords{
astrometry --
Galaxy: kinematics and dynamics --
masers --
stars: formation --
stars: individual: W43 --
techniques: high angular resolution
}

\maketitle

\section{INTRODUCTION}
\label{sec:intro}

\begin{figure}[H]
  \begin{center}
    \includegraphics[scale=0.60,angle=-0]{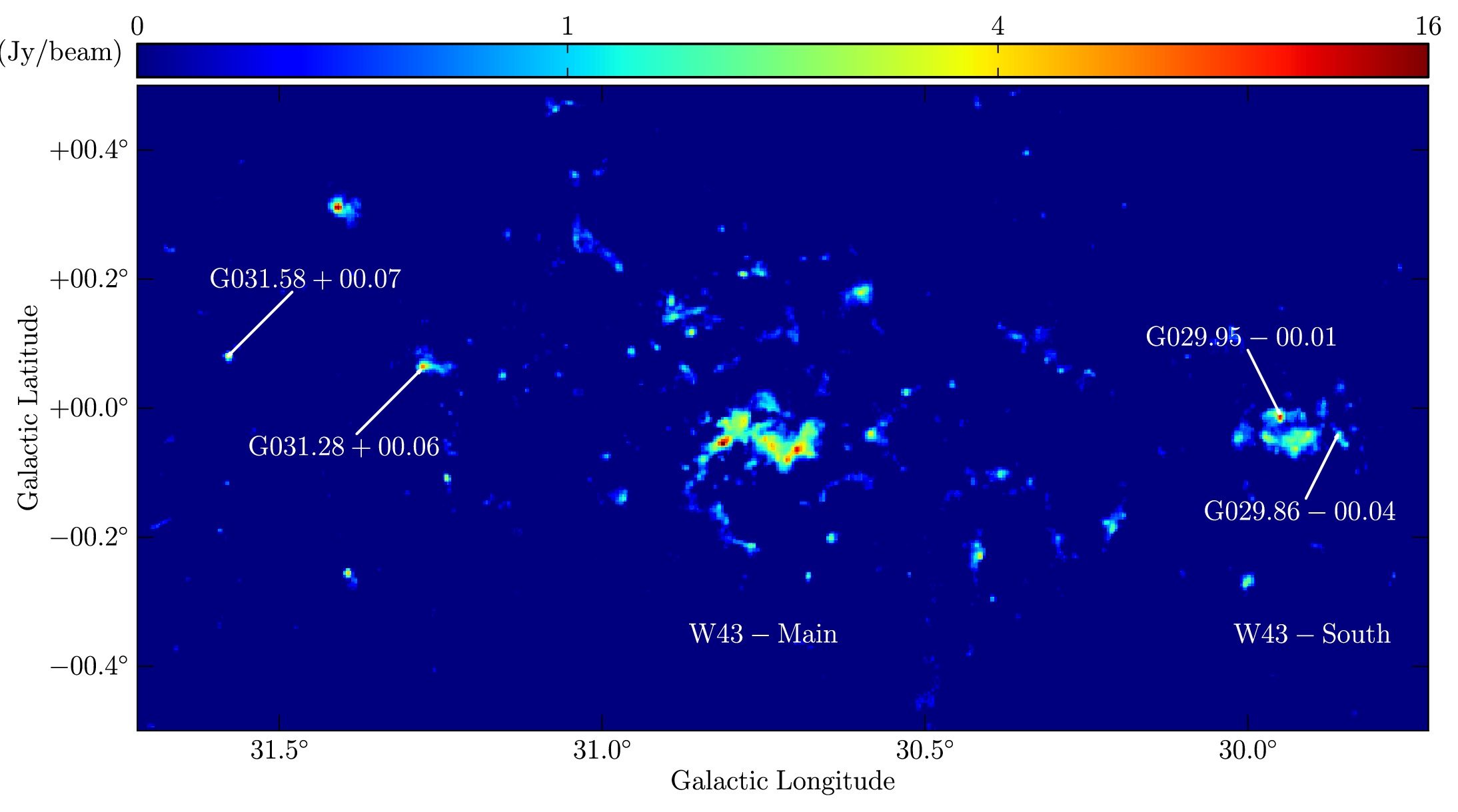}
  \end{center}
  \caption{
  870 $\mu$m dust continuum emission of the W43 complex 
  from the APEX Telescope Large Area Survey of the Galaxy
  (ATLASGAL,~\citealt{2009A&A...504..415S}).\newline
  (A color version of this figure is
  available in the online journal.)
  \label{fig:w43}}
\end{figure}

W43 is one of the most prominent regions of massive star formation in
the inner Galaxy~\citep{2011A&A...529A..41N}. As shown in
Figure~\ref{fig:w43}, it contains two of the largest molecular cloud
groups of the first Galactic quadrant: W43-Main at $\ell~\approx$
30\decdeg7 and W43-South at $\ell~\approx$ 29\decdeg9.  W43-Main is
famous for its giant \HII\ region powered by a cluster of OB and
Wolf-Rayet stars, which emit a Lyman continuum luminosity of $10^{51}$
ionizing photons per second, and has a far-infrared luminosity of 3.5
$\times~10^6$ \Lsun\
\citep{1978A&A....66...65S,1985ApJ...296..565L,1999AJ....117.1392B}.
For the very active star-forming phenomena in its center, W43-Main is
considered a ``mini-starburst'', and three extremely massive dense cores
W43-MM1, W43-MM2 and W43-MM3 were suggested to be potentially forming
high-mass stars by \citet{2003ApJ...582..277M} and confirmed by
\citet{2010A&A...518L..90B}.
Recently, \citet{2013arXiv1306.0547N} discovered two ridge-like
structures and extended SiO emission toward W43-Main; the spatial and
velocity overlap between the high-density ridge and the shocked SiO gas
suggest that the ridges could be forming via colliding flows driven by
gravity and accompanied by low-velocity shocks.

Accurate distances are crucial to derive fundamental physical parameters
of the molecular clouds. Size depends directly on distance and
luminosity scales as the square of distance.  The commonly used methods
to determine the distance of W43 are kinematic distance estimates based
on radial velocity measurements and a Galactic rotation curve, and
luminosity distances based on dereddened colors and spectral type.
However, kinematic distance have uncertainties, often $>$ 20\%, which
are highly dependent on location in the
Galaxy~\citep{2009ApJ...700..137R}; for cases with large peculiar
motions can be over a factor of two~\citep{2006Sci...311...54X}.  In
addition, one needs to add uncertainties from the adopted values of
Galactic parameters whereas trigonometric parallax is essentially a
direct and assumption free method.  Due to difficulties in classifying
spectral types and calibrating luminosities, it is hard to determine
distances accurately by photometry and
spectroscopy~\citep[e.g.,][]{1999AJ....117.1392B}.   Previous
investigators have derived distances of 4 -- 7 kpc to W43
~\citep[e.g.,][]{1978A&A....66...65S,1999AJ....117.1392B,2011MNRAS.411..705M}.

The most reliable method for astronomical distance measurement is the
trigonometric parallax.  Recently, trigonometric parallax measurements
of masers in star-forming regions using the Very Long Baseline
Interferometry (VLBI) techniques have shown that the parallaxes can be
accurately determined at the level of $\sim10$ \uas~\citep[and
references therein]{2009ApJ...700..137R}.  Motivated by these
measurements, we are using the NRAO\footnote{The National Radio
Astronomy Observatory (NRAO) is a facility of the National Science
Foundation operated under cooperative agreement by Associated
Universities, Inc} Very Long Baseline Array (VLBA) to conduct a key
science project, the Bar and Spiral Structure Legacy (BeSSeL)
Survey\footnote{\url{http://bessel.vlbi-astrometry.org/}}
~\citep{2011AN....332..461B}, to study the structure and kinematics of
the Galaxy by measuring parallaxes and proper motions for hundreds of 22
GHz \hho\ and 6.7/12.2 GHz \meth\ maser sources associated with massive
star-forming regions.

Several masers have been discovered in W43 (within the region 29\deg\ $
< \ell < 32$\deg, $-1$\deg\ $< b < +$1\deg).  Figure~\ref{fig:w43}
shows two source pairs in W43: \GIXa\ and \GIXb\ located in W43-South
and \GIa\ and \GIb\ located in the northern portion of W43-Main.
\GIXa\ and \GIXb\ are ultra-compact (UC) \HII\ regions powered by an O
star. The 6.7 GHz~\citep{1991ApJ...380L..75M} and 12.2
GHz~\citep{1995MNRAS.274.1126C} \meth\ maser sites and Local Standard of
Rest (LSR) velocities coincide with values determined for \hho\
masers~\citep{1996A&AS..120..283H} and an NH$_3$ hot core, which is
believed to be a signpost of an embedded high-mass star in an early
stage of formation~\citep{1998A&A...331..709C}.
\GIa\ is an UC \HII\ region and a candidate for hosting a massive
protostar~\citep{2000A&A...362.1093M, 2001A&A...369..278M,
2002A&A...387..179M}. The 12 GHz methanol maser in \GIa\ was discovered
by \citet{1988ApJ...326..931K}.  Based on a multi-wavelength study, this
methanol maser site was found to be associated with an \HII\ region and
an Infrared Astronomical Satellite (IRAS) source, but it is separated
from the radio peak by 3.5 $\times$ 10$^4$ AU, assuming a distance of
5.6 kpc~\citep{2005A&A...429..945M}.
\GIb\ hosts a \hho\ maser and a 6.7 GHz \meth\ maser
source~\citep{2000A&AS..143..269S}, and it is associated with an \HII\
region~\citep{2009ApJS..181..255A} and the IRAS source 18461$-$0113
~\citep{1996A&AS..115...81B}.

In this paper, we present results of parallax and proper motion
measurements of the four masers \GIXa, \GIXb, \GIa, and \GIb\ in the
massive star-forming complex W43 from VLBA observations that are part of
the BeSSeL Survey.  These sources are located in the first quadrant of
the Galaxy, where kinematic distances are subject to a kinematic
distance ambiguity.  Our direct measurements of distance solve the
ambiguity for W43 and help locate the spiral arm in which W43 resides.
This star-forming complex is located toward the end of the central bar
of the Galaxy.  Combining distance, \VLSR, and proper motion of the
masers yields their full space motions, which help us to study the
kinematics of this complicated region.

\section{OBSERVATIONS AND CALIBRATION PROCEDURES}

Our observations of 12 GHz methanol and 22 GHz water masers, together
with several compact extragalactic radio sources, were carried out under
VLBA programs BR145O, BR145Q, and BR145W and spanned one year.  For
these sources, the parallax signature in Declination is considerably
smaller than for Right Ascension, and we scheduled the observations so
as to maximize the Right Ascension parallax offsets as well as to
minimize correlations among the parallax and proper motion parameters.

Our general observing setup and calibration procedures are described in
\citet{2009ApJ...693..397R}; here, we discuss only aspects of the
observations that are specific to the maser sources presented in this
paper.  We used four adjacent intermediate frequency (IF) bands of 8 MHz
and recorded both right and left circular polarizations (RCP and LCP).
The second IF band was centered on the maser signal at an \VLSR\
velocity of 100 \kms\ for \GIXa, 96 \kms\ for \GIXb, 110 \kms\ for \GIa,
and 99 \kms\ for \GIb.  We observed three International Celestial
Reference Frame (ICRF) sources~\citep{1998AJ....116..516M}, near the
beginning, middle, and end of the phase-referencing observations to
monitor delay and electronic phase differences among the observing
bands.  The observing frequency, velocity resolution for line
observation, and epochs are listed in Table~\ref{tab:obs}.

At each epoch, the observations consisted of four 0.5-hour ``geodetic
blocks'' (used to calibrate and remove unmodeled atmospheric signal
delays), and three 1.7-hour periods of phase-referenced observations
inserted between these blocks.  In the phase-referenced observations, we
cycled between the target maser and several background sources,
switching sources every 30 to 40 seconds.  Table~\ref{tab:src} lists the
observed source positions, intensities, source separations, reference
maser LSR velocities, and restoring beams.

The interferometric correlation was performed with the
DiFX\footnote{DiFX: A software Correlator for VLBI using Multiprocessor
Computing Environments, is developed as part of the Australian Major
National Research Facilities Programme by the Swinburne University of
Technology and operated under licence} software correlator
\citep{2007PASP..119..318D} in Socorro, NM.  The data reduction was
conducted using the NRAO's Astronomical Image Processing System (AIPS)
together with scripts written in ParselTongue
\citep{2006ASPC..351..497K}.  A spectral channel with strong and compact
maser emission was used as the interferometer phase reference.  After
performing the calibration for the polarized bands separately, we
combined the RCP and LCP bands to form Stokes I and imaged the continuum
emission of the background sources, integrating the four IF bands using
the AIPS task IMAGR.  For the masers, we also formed Stokes I data and
then imaged the emission in each spectral channel.

We determined source positions by fitting elliptical Gaussian brightness
distributions to the images of strong maser spots and background sources
using the AIPS task SAD or JMFIT.
As discussed by \citet{2010ApJ...720.1055S}, relative positions for
sources (i.e., maser and quasar sources) at low antenna elevations are
very sensitive to atmospheric delay errors
\citep[e.g.][]{2008PASJ...60..951H}.  Further, our sources are located
at near-zero Declination, where phase calibration errors can lead to
limited dynamic range images.   Therefore, we adopted an elevation
cutoff of 25\deg\ -- 30\deg, below which data were discarded for each
antenna, allowing for improved images with dynamic range of better than
15:1.

\section{ASTROMETRIC PROCEDURES AND RESULTS}
\label{sec:rst}

We considered maser spots at different epochs to be persistent if their
positions in the same spectral channel were coincident within 3 \masy\
$\times$ $\Delta t$ yr, where $\Delta t$ is the time gap between two
epochs; this limit corresponds to a linear motion of less than 85 \kms\
at a distance of 6 kpc.

Data used for parallax and proper motion fits were residual position
differences between maser spots and background sources in eastward (\dx\
= $\Delta\alpha\cos\delta$) and northward (\dy\ = $\Delta\delta$)
directions.  The data were modeled as a combination of the parallax
sinusoid (determined by a single parameter, the parallax) and a linear
proper motion for each coordinate.  Because systematic errors (owing to
small uncompensated atmospheric delays and, in some cases, varying maser
and calibrator source structures) typically dominate over thermal noise
when measuring relative source positions, we added ``error floors'' in
quadrature to the formal position uncertainties (see values listed in
Tables~\ref{tab:g029a_para_pm} -- \ref{tab:g031b_para_pm}). We used
independent error floors for the \dx\ and \dy\ data and adjusted them to
yield post-fit residuals with reduced $\chi_\nu^2$ near unity for both
coordinates.

\begin{figure}[H]
  \centering
  \includegraphics[angle=-0,scale=0.60]{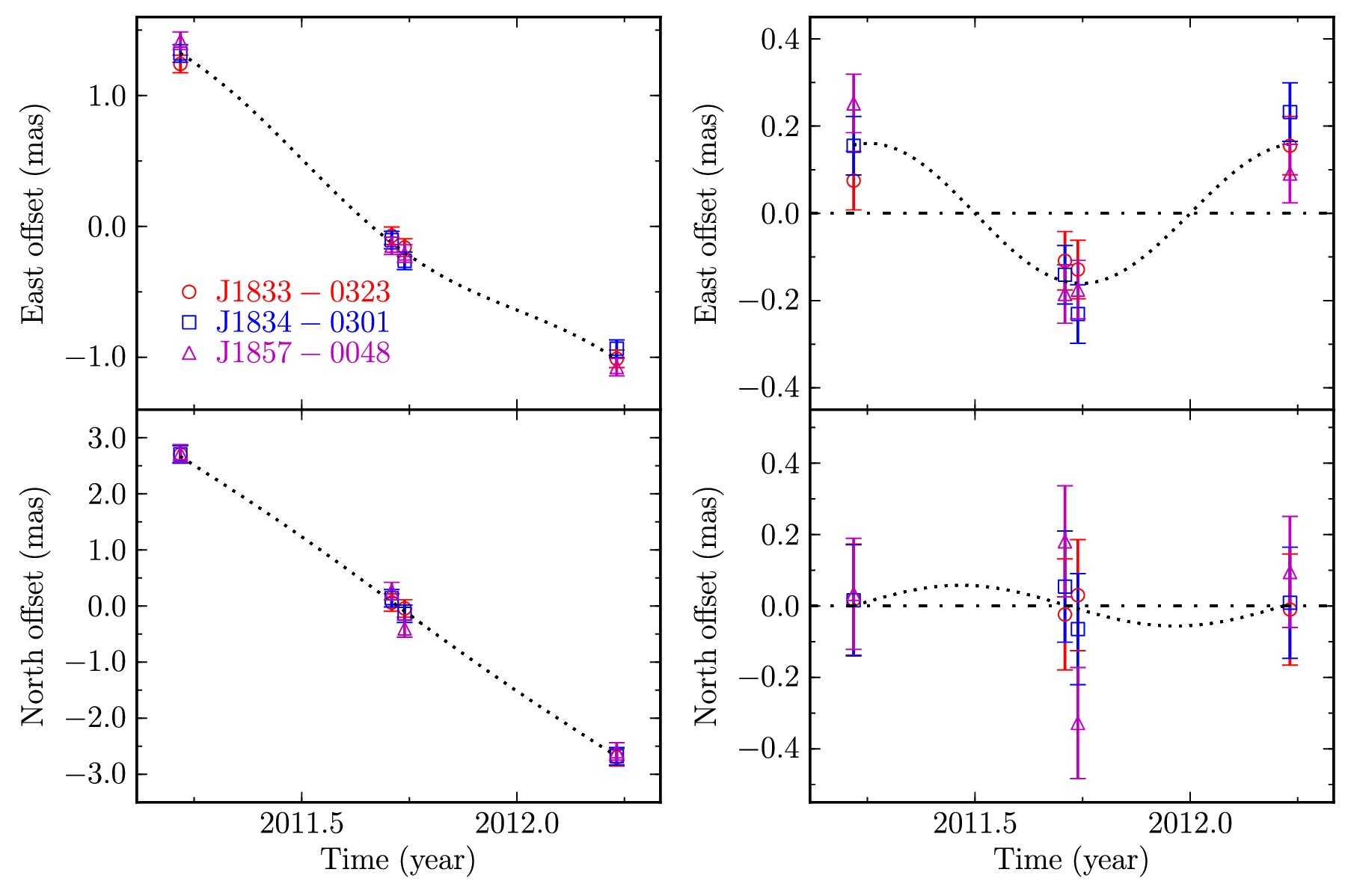}
  \caption{
  Parallax and proper motion data ({\it markers}) and best-fitting model
  ({\it dotted lines}) using the reference maser spot at \VLSR\ of 100.0
  \kms\ as an example for \GIXa.   Plotted are positions of the maser spot
  relative to the extragalactic radio sources \JIII\ ({\it circles}),
  \Jfour\ ({\it squares}), and \Jseven\ ({\it triangles}).  {\it Left
  panel}: Eastward ({\it upper panel}) and northward ({\it lower panel})
  offsets versus time.  {\it Right panel}: Same as the {\it left panel},
  except the best-fitting proper motion has been removed, displaying only
  the parallax signature.\newline (A color version of this figure is available in
  the online journal.)
  \label{fig:g029a_para}}
\end{figure}

\begin{figure}[H]
  \centering
  \includegraphics[angle=-0,scale=0.60]{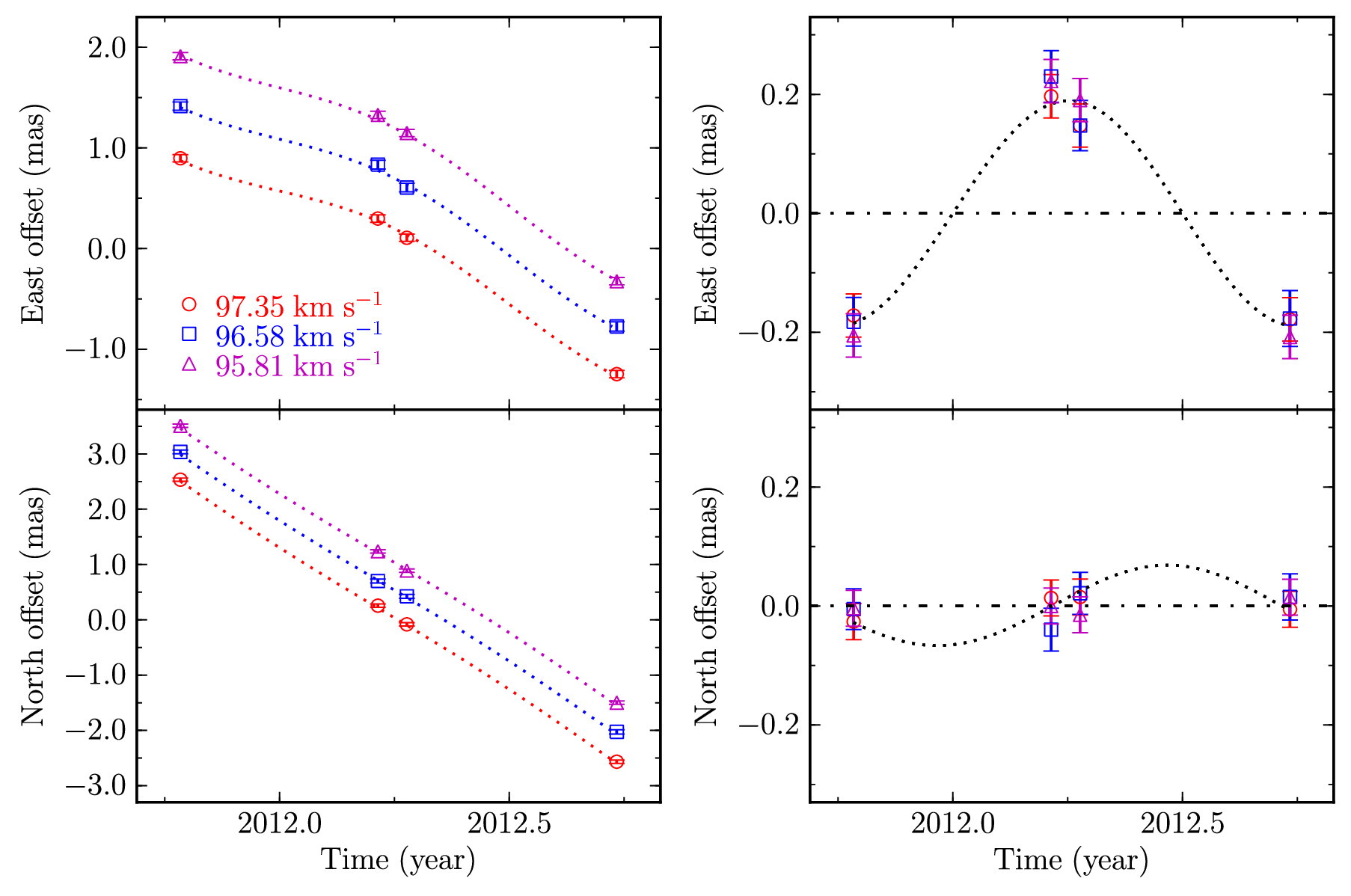}
  \caption{
  Parallax and proper motion data ({\it markers}) and best-fitting model
  ({\it dotted lines}) for \GIXb.  Plotted are positions of the maser spots
  (denoted with different marks) relative to the extragalactic radio
  source \Jfour.  {\it Left panel}: Eastward ({\it upper panel}) and
  northward ({\it lower panel}) offsets versus time.  {\it Right panel}:
  Same as the {\it left panel}, except the best-fitting proper motion has
  been removed, displaying only the parallax signature. (A color version
  of this figure is available in the online journal.)
  \label{fig:g029b_para}}
\end{figure}

\begin{figure}[H]
  \centering
  \includegraphics[angle=-0,scale=0.60]{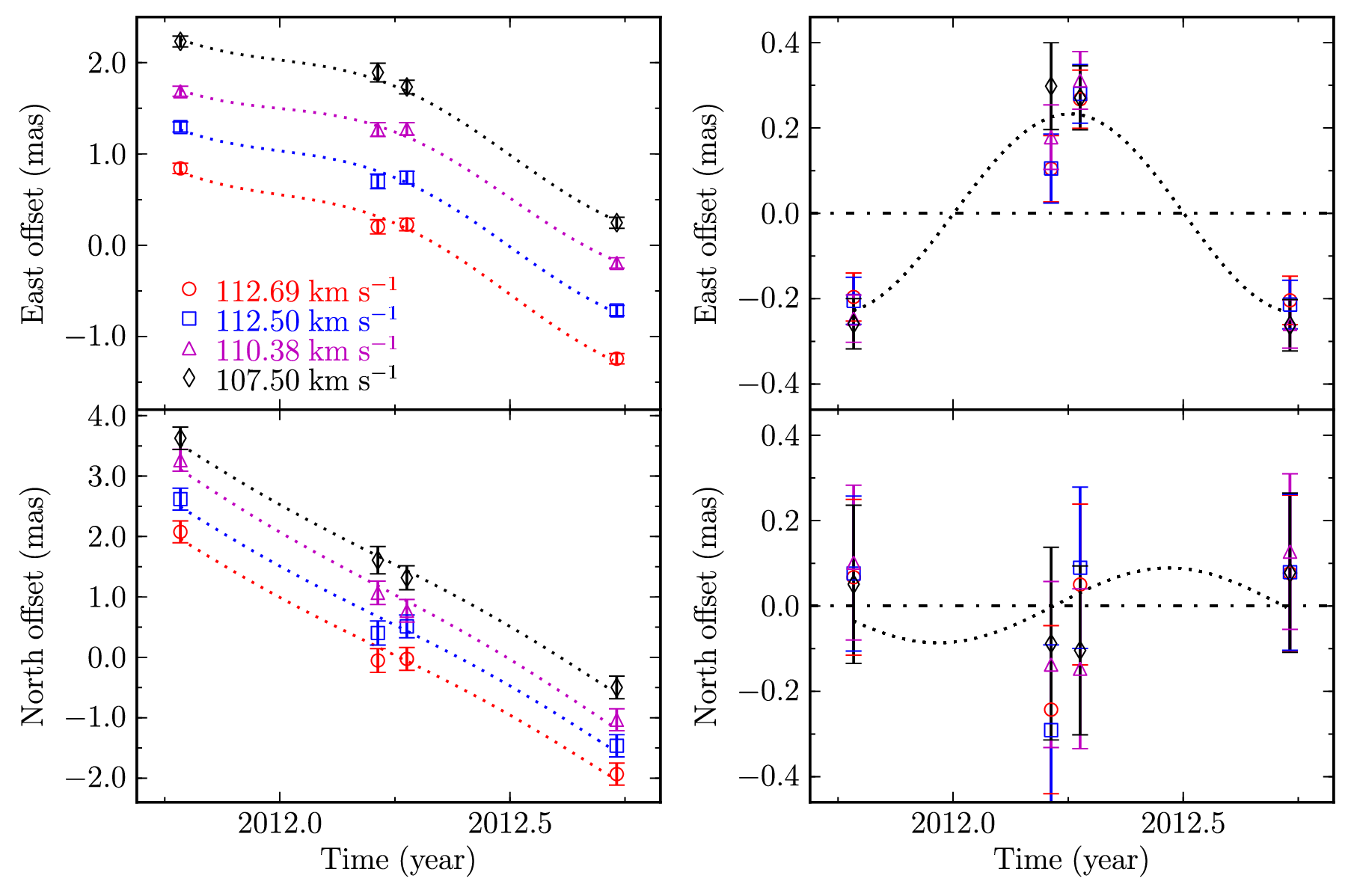}
  \caption{
  Parallax and proper motion data ({\it markers}) and best-fitting model
  ({\it dotted lines}) for \GIa.  Plotted are positions of the maser spots
  (denoted with different marks) relative to the extragalactic radio
  source \Jthree.  {\it Left panel}: Eastward ({\it upper panel}) and
  northward ({\it lower panel}) offsets versus time.  {\it Right panel}:
  Same as the {\it left panel}, except the best-fitting proper motion has
  been removed, displaying only the parallax signature. (A color version
  of this figure is available in the online journal.)
  \label{fig:g031a_para}}
\end{figure}

\begin{figure}[H]
  \centering
  \includegraphics[angle=-0,scale=0.60]{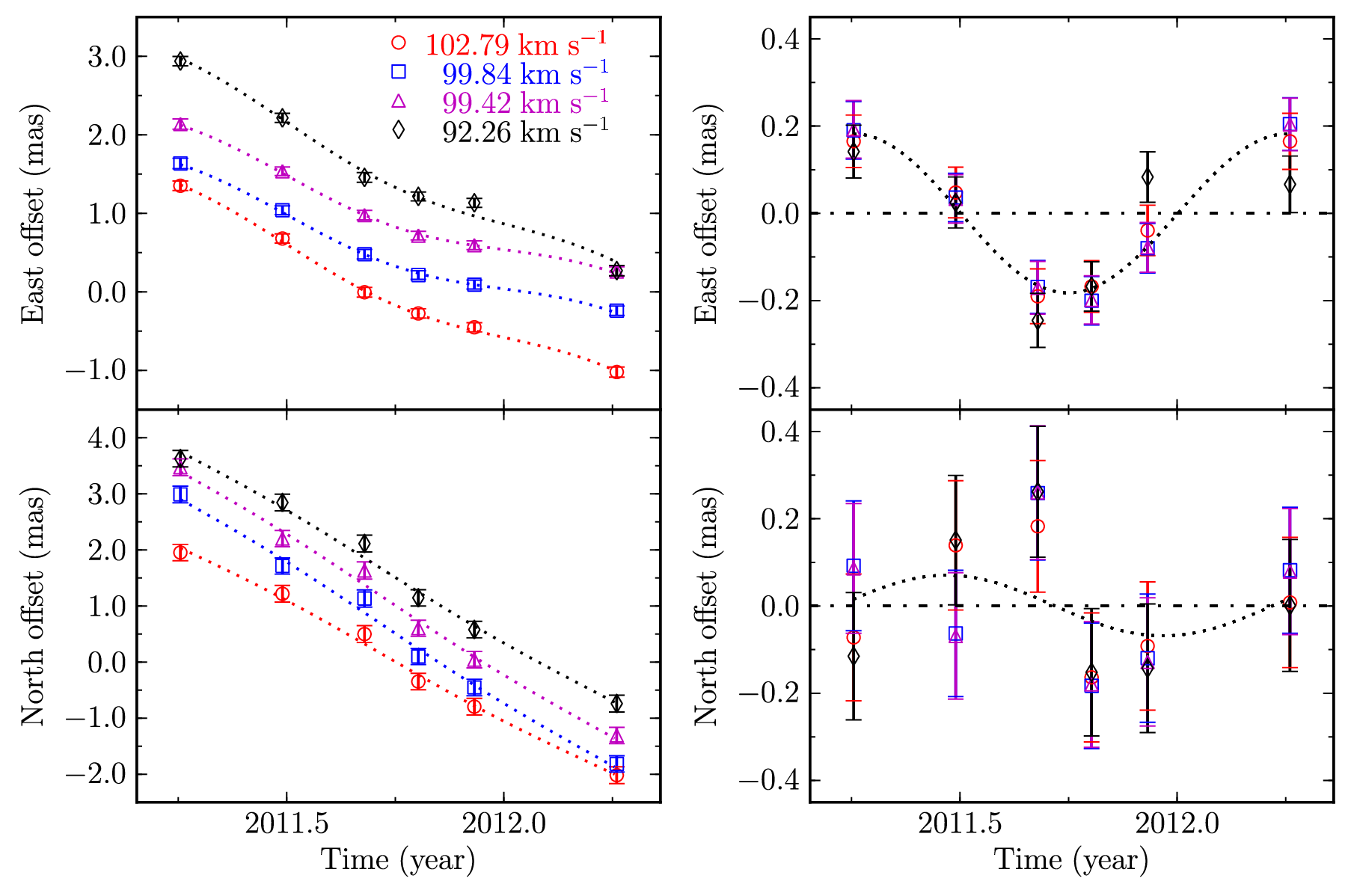}
  \caption{
  Parallax and proper motion data ({\it markers}) and best-fitting model
  ({\it dotted lines}) for \GIb.  Plotted are positions of the maser spots
  (denoted with different marks) relative to the extragalactic radio
  source \Jthree.  {\it Left panel}: Eastward ({\it upper panel}) and
  northward ({\it lower panel}) offsets versus time.  {\it Right panel}:
  Same as the {\it left panel}, except the best-fitting proper motion has
  been removed, displaying only the parallax signature. (A color version
  of this figure is available in the online journal.)
  \label{fig:g031b_para}}
\end{figure}

The apparent motions of the maser spots can be altered by a combination
of spectral blending and changes in intensity. Thus, for parallax
fitting, one needs to find stable, unblended spots and preferably use
many maser spots to average out any residual effects.  To achieve this
goal, we first fitted the position offsets of each maser spot relative
to each background source separately and excluded those maser spots with
large parallax uncertainties ($>$ 40\%).  Since one expects that the
parallaxes of all maser spots are identical within measurement
uncertainties (while the proper motions can vary owing to internal
motions), we did a combined solution fitting a single parallax to all
maser spots and background sources, but allowing for different proper
motions of the maser spots.  The quoted parallax uncertainty is the
formal error multiplied by $\sqrt{N}$, where N is the number of maser
spots fitted, in order to allow for the possibility of correlated
position errors of the maser spots relative to the extragalactic
continuum sources.  The position data and the parallax fits are shown in
Figures~\ref{fig:g029a_para} -- \ref{fig:g031b_para}.

Since all maser sources in our observations have simple spectra which
span $\le$ 10 \kms, and the proper motions of maser spots (relative to
common or different background sources) are consistent with each other
within $\pm2\sigma$ (suggesting that the internal motions are small), we
estimated the absolute proper motion of the central star exciting the
masers by taking the variance-weighted mean of the proper motions.  The
uncertainty of the mean proper motion is evaluated using standard error
propagation.

Details of the measurement of parallaxes and proper motions of the
individual sources are presented in the \nameref{sec:app}.  The \VLSR\
of each source is estimated using the median value of the maser emission
velocities , which is consistent with that derived from the Gaussian
fits of the Galactic Ring Survey (GRS, \citealt{2006ApJS..163..145J})
\coIII\ spectrum.

Based on differences between the \VLSR\ of the masers and that
indicated by thermal CO emission, we assign a \VLSR\ uncertainty of 3
and 5 \kms\ for \meth\ and \hho\ maser sources, respectively.   A
similar uncertainty is associated with our estimate of the proper motion
of the central star, and we added in quadrature a proper motion
uncertainty of 5 \kms\ (at the measured distance) to the measurement
uncertainties presented in the \nameref{sec:app}.  The measured
quantities and fitted parameters and their uncertainties are summarized
in Table~\ref{tab:ppm}.

\section{DISCUSSION}

\subsection{Galactic arm assignment of W43}

\begin{figure}[H]
  \begin{center}
    \includegraphics[scale=0.51,angle=-0]{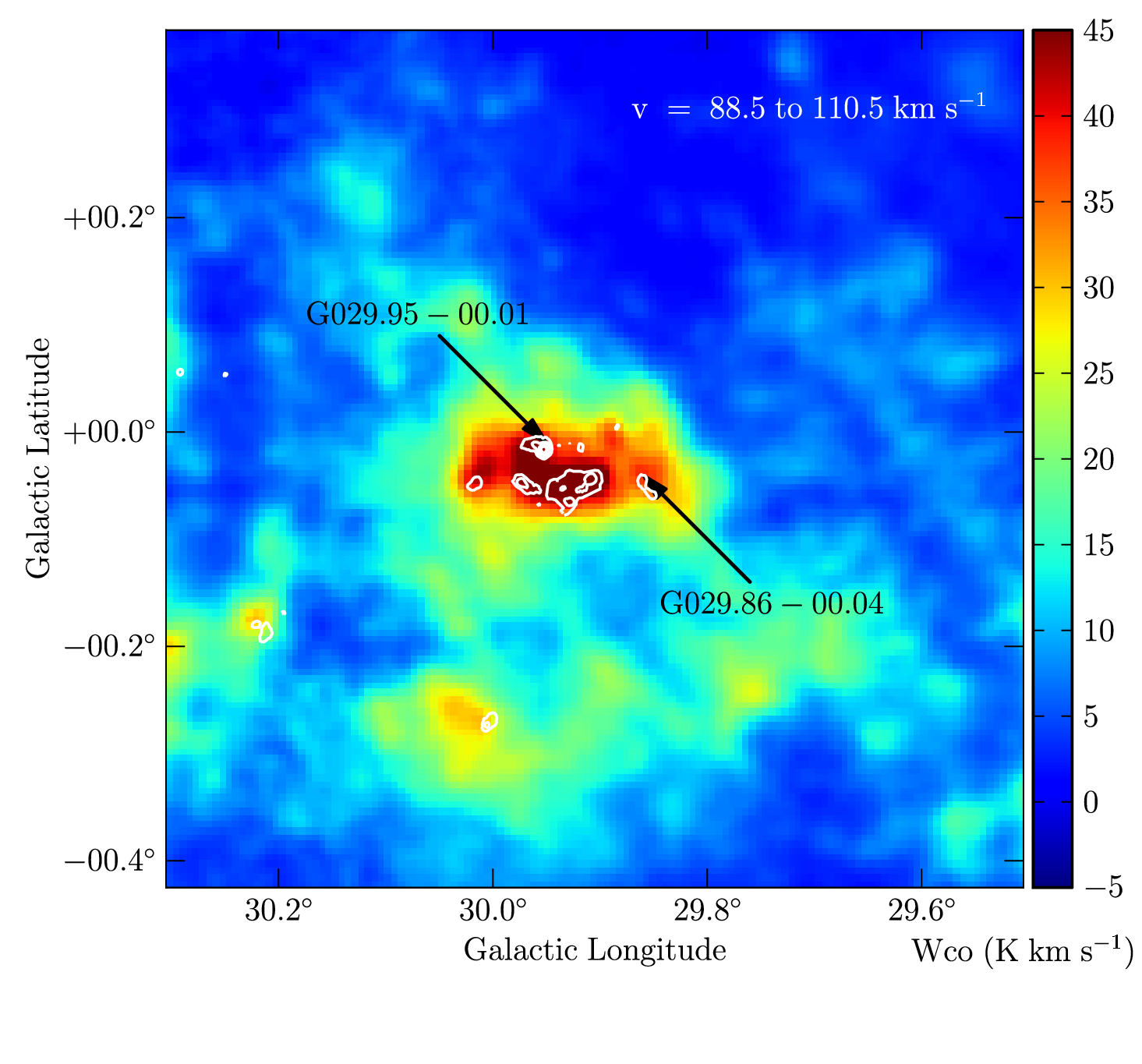}
    \hspace{1mm}
    \includegraphics[scale=0.51,angle=-0]{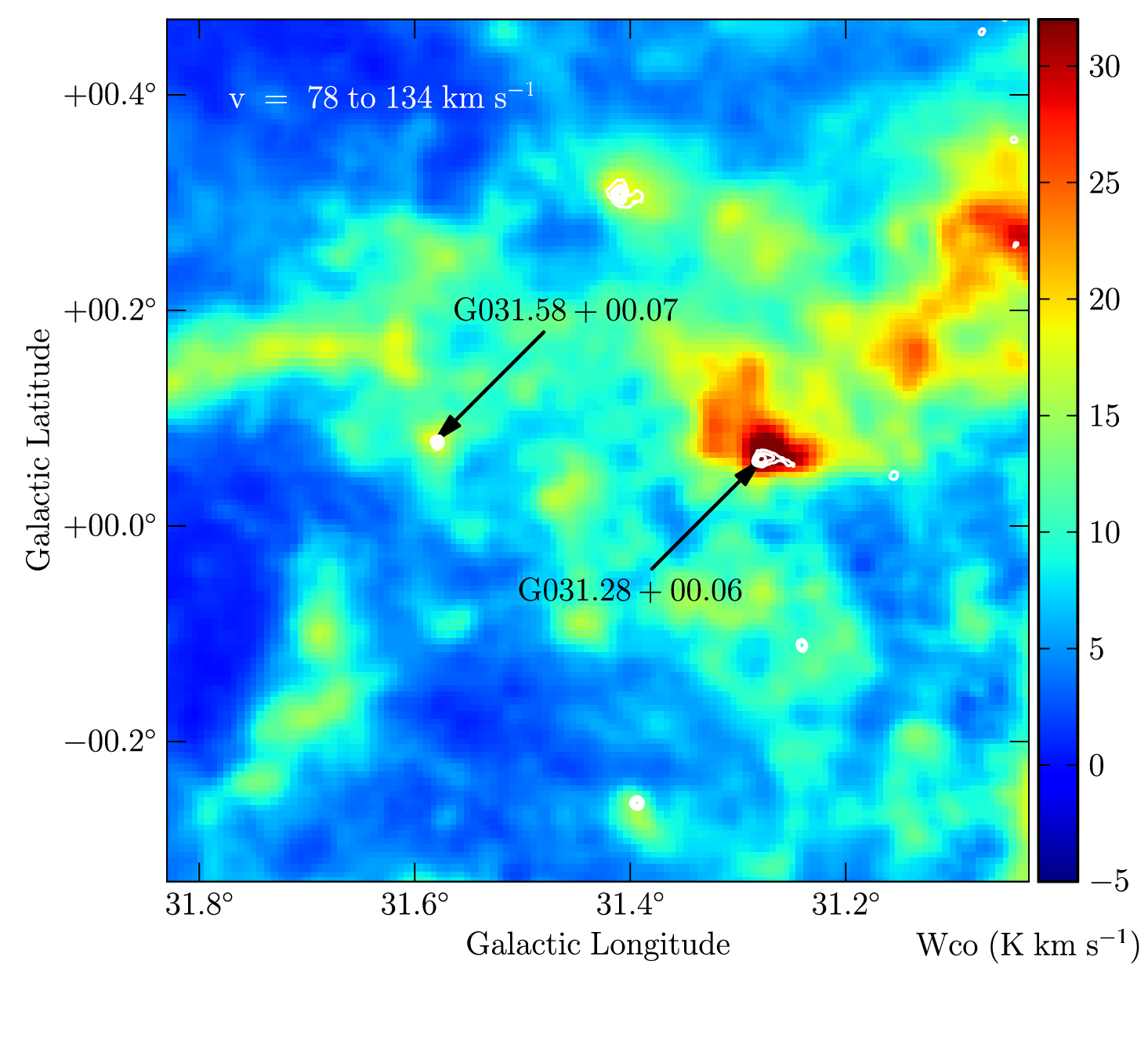}
  \end{center}
  \caption{
  Velocity-integrated \coIII\ intensity of the molecular clouds
  apparently associated with \GIXa\ and \GIXb\ ({\it left panel}), and
  with \GIa\ and \GIb\ ({\it right panel}). The ranges of velocity
  integration are indicated in the top of each figure.  The data are
  from the Galactic Ring Survey~\citep{2006ApJS..163..145J}. Over
  plotted contours show the 870 $\mu$m continuum emission from the APEX
  Telescope Large Area Survey of the Galaxy
  (ATLASGAL,~\citealt{2009A&A...504..415S}). Contour levels start from 1
  \jybeam\ and increase by a factor of 2.
  (A color version of this figure is available in the online journal.)
  \label{fig:wco}}
\end{figure}

Spiral arms in the Galaxy can be identified as large-scale features in
longitude-velocity ($\ell-v$) diagrams from \HI\
\citep[e.g.][]{1972A&A....16..118S} and CO surveys
\citep[e.g.][]{1980ApJ...239L..53C,1986ApJ...305..892D}.  Assigning our
masers to arms by associating them with molecular clouds avoids the bias
of the alternative method of using the maser distance and a model of
spiral structure.  A secure arm assignment requires that the position
and velocity of the maser and the molecular cloud be in agreement with
the position-velocity ($\ell-v$) locus of the arm.  Using the data from
the \coIII\ GRS by \citet{2006ApJS..163..145J} and the
APEX\footnote{APEX, the Atacama Pathfinder EXperiments, is a
collaboration between the Max Planck Institut f\"ur Raodioastronomie,
the Onsala Space Observatory, and the European Southern Observatory.}
Telescope Large Area Survey of the GALaxy (ATLASGAL,
\citealt{2009A&A...504..415S}), we find that all the four sources have
good position-velocity correspondence with molecular clouds (see
Figure~\ref{fig:wco}).

\begin{figure}[H]
  \begin{center}
    \includegraphics[scale=0.90]{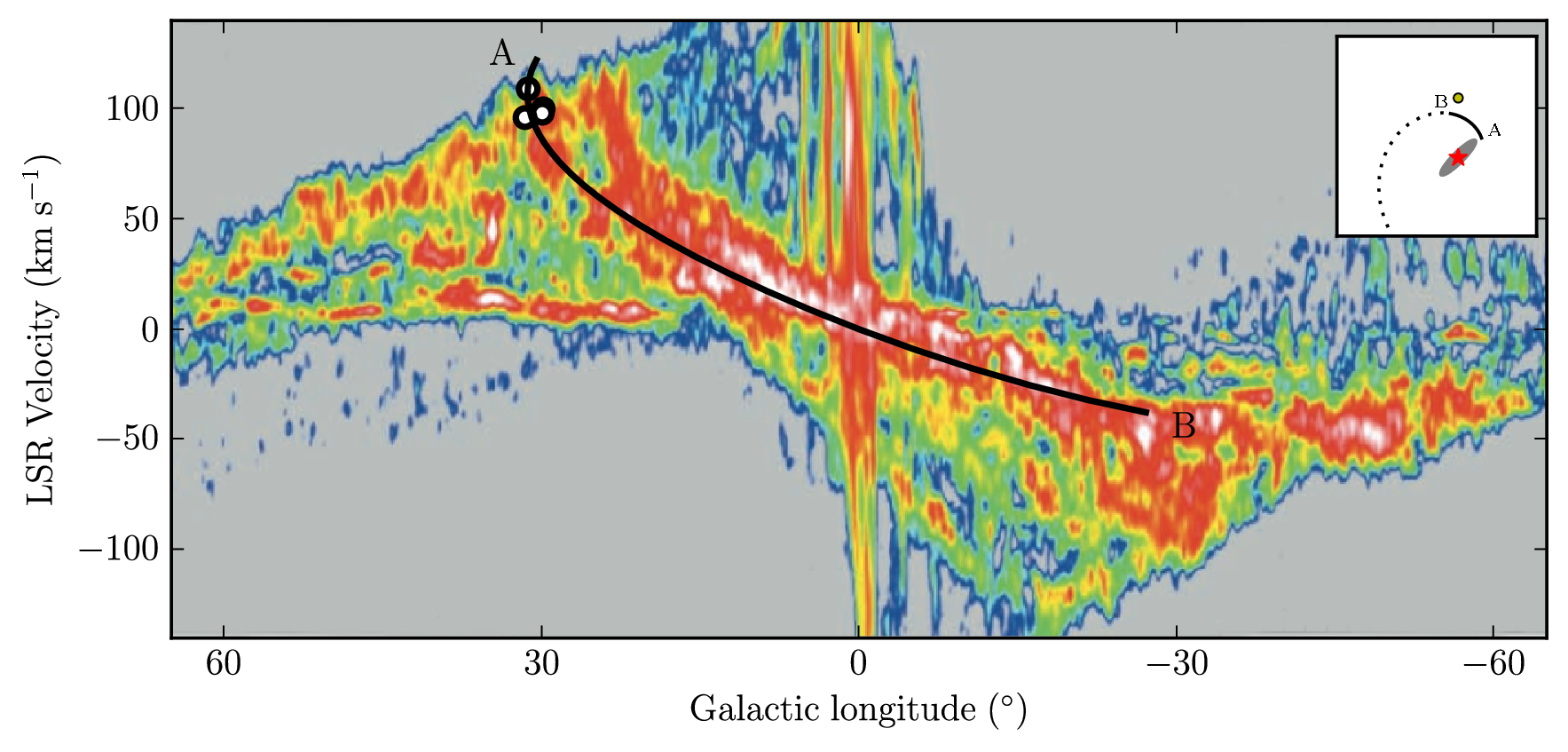}
  \end{center}
  \caption{
  CO longitude-velocity diagram adapted from \citet[][their figure
  3]{2001ApJ...547..792D}.  CO intensities integrated over a strip
  $\sim$ 4\deg\ wide in Galactic latitude centered on the Galactic plane
  are displayed logarithmically from 0.1 K arcdeg (gray) to 10 K arcdeg
  (white).  The solid curve line outlines the quasi-continuous
  structures in $\ell-v$ which are assumed be a portion of the Scutum
  arm~\citep[e.g.][]{1972A&A....16..118S, 1980ApJ...239L..53C,
  2011ApJ...734L..24D}. The circles mark the $\ell-v$ positions of the
  four maser sources in W43.
  The inset is a schematic face-on view of the Galaxy showing only the
  Scutum-Centaurus spiral arm (solid and dotted line). The gray eclipse
  indicates a long thin bar with a semi-major axis of 3.4 kpc and an
  orientation of 45\deg\ from \citet{2011ApJ...733...27G}. The circle
  and star mark the positions of the Sun and the Galactic Center,
  respectively.  The letters A and B in both the main figure and the
  inset mark the Scutum arm loci at $\ell$ = 31\decdeg5 and $-$27\deg,
  respectively.
  (A color version of this figure is available in the online journal.)
  \label{fig:lv}}
\end{figure}

The clouds associated with the four maser sources in the W43 complex
have angular sizes of $\approx$ 3 -- 12\arcmin, relatively large
composite line widths (FWHM $\approx$ 10 \kms), and high positive LSR
velocities of $\approx$ 95 -- 110 \kms, all of which are consistent with
the $\ell-v$ locus of the Scutum spiral arm shown in
Figure~\ref{fig:lv}.  This indicates that the four maser sources in W43
can be confidently located in the Scutum spiral arm.

\subsection{Distance to W43}

At the kinematic distance of 6 kpc from the Sun, W43 is assumed to be
located near the meeting point of the Scutum arm and the bar in the
Galactic plane~\citep{2011A&A...529A..41N}.  A solid distance
measurement to this region will significantly contribute to our
understanding of the physical properties of this star forming complex
and the structure of the Milky Way.

As listed in Table~\ref{tab:ppm}, the distances and proper motions of
the source pairs, \GIXa\ and \GIXb, and \GIa\ and \GIb, are consistent
within their joint uncertainties. This suggests that each source pair is
located at the same distance.  The variance-weighted average parallax is
0.175 $\pm$ 0.014 mas (5.71$^{+0.50}_{-0.42}$ kpc) and 0.203 $\pm$ 0.025
mas (4.93$^{+0.69}_{-0.54}$ kpc) for the former and latter source pairs,
respectively.
Since the two source pairs share similar positions and \VLSR, and have
approximately equal distance (within the uncertainties), it is very
likely that they belong to the W43 cloud complex.  Thus, we estimated a
grand average parallax of 0.182 $\pm$ 0.012 mas (5.49$^{+0.39}_{-0.34}$
kpc) to W43 using a variance-weighted average of the four sources.
Our measured distance to W43 is in good agreement with the latest
spectroscopic distance of 5.2 $\pm$ 0.5 kpc by C-H.~R. Chen et al.
(2013, in preparation), using three O stars.

Our measured parallax distance of 5.49 kpc to W43 is about 90\% of the
previously adopted near kinematic distance of 6 kpc (e.g.,
\citealt{2011A&A...529A..41N}).  This implies that the mass and
luminosity of the molecular complex W43 have been overestimated in
\citet{2011A&A...529A..41N} by factor of 1.2 $\pm$ 0.14.  The revised
total H$_2$ mass from \coII\ and cloud mass from \coIII\ are 5.9
$\times~10^6$ and 3.5 $\times~10^6$ \Msun, and the 8 $\mu$m luminosity
becomes 1.32 $\times~10^7$ \Lsun.

\subsection{Galactic locations and peculiar motions}

\begin{figure}[H]
  \centering
  \includegraphics[angle=-0,scale=0.80]{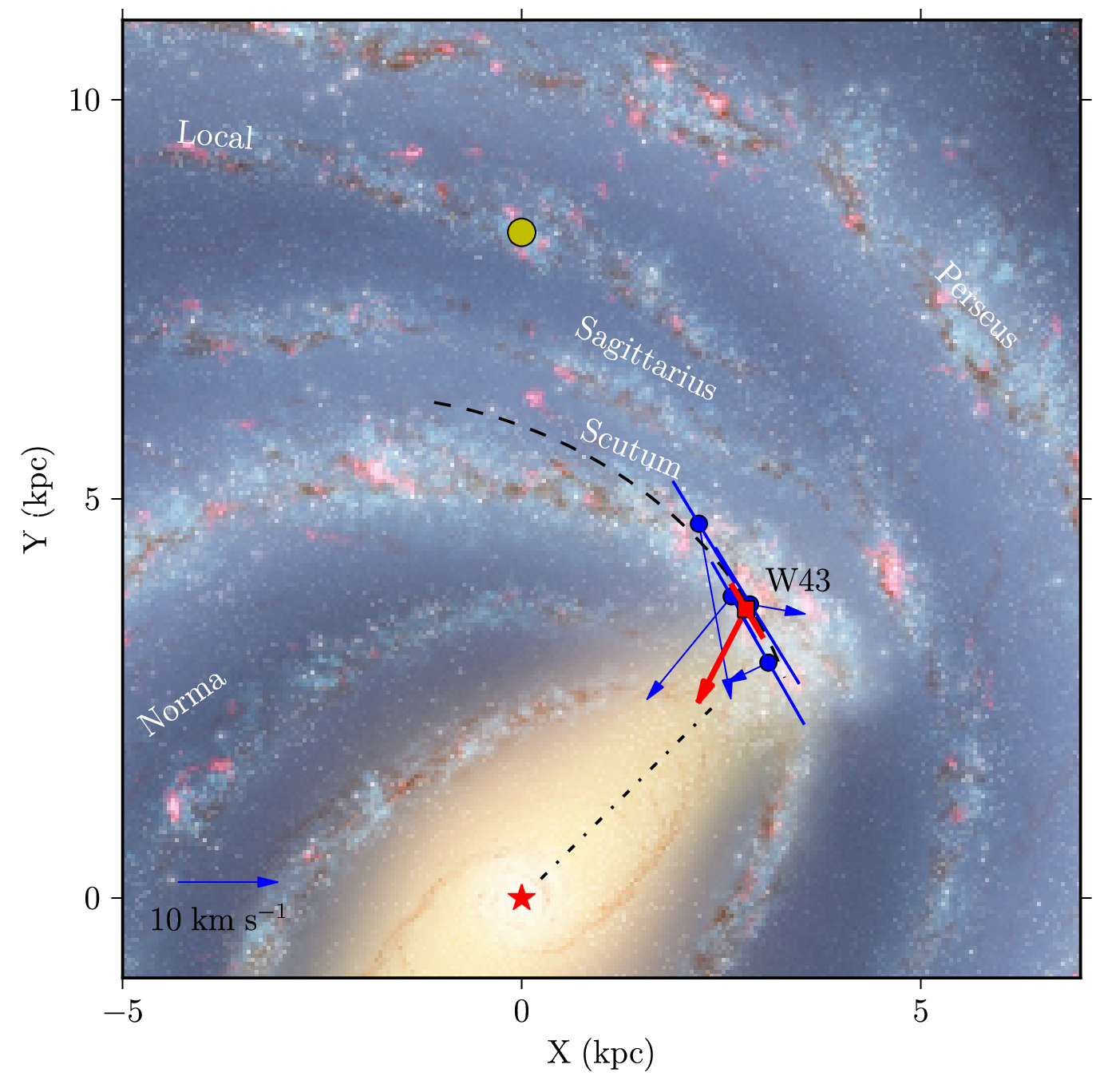}
  \caption{
  Locations ({\it red square for W43 and blue dots for the others}) and
  peculiar motions ({\it red thick arrow for W43 and blue thin arrows
  for the others}) of maser sources in W43 projected on the Galactic
  plane. The values of location and peculiar motion for W43 are
  calculated using the parameters listed in Table~\ref{tab:ppm}, which
  are averaged from the other four sources.  The length of the bar
  denotes the distance uncertainty. A 10 \kms\ motion scale is in the
  lower left.  The background is an artist's conception of the Milky Way
  (R. Hurt: NASA/JPL-Caltech/SSC) viewed from the north Galactic pole
  from which the Galaxy rotates clockwise.  The Galactic Center ({\it
  red asterisk}) is at (0, 0) and the Sun ({\it yellow dot}) at (0,
  8.34) kpc.  The dashed curve line denotes the Scutum arm position with
  a pitch angle of 20\decdeg1 from \citet{2013Sato}.  The
  dash-dotted line starting at the Galactic Center denotes a long thin
  bar with a semi-major axis of 3.4 kpc and an orientation of 45\deg\
  from \citet{2011ApJ...733...27G}.  (A color version of this figure is
  available in the online journal.)
  \label{fig:pec_mot}}
\end{figure}

Combining the four sources reported here and the twelve sources with
parallax measurements in the Scutum arm from the BeSSeL survey and the
Japanese VLBI Exploration of Radio Astrometry (VERA) project, makes a
total of 16 sources in the Scutum arm.  These well delineate a section
of the arm with a pitch angle of 20\decdeg1~\citep{2013Sato}.  Based on
the 6.7 GHz \meth\ maser population, \citet{2011ApJ...733...27G}
suggested that the inner region of the Galaxy contains a thin bar with a
length of 3.4 kpc toward a Galactocentric azimuth of $\approx$ 45\deg\ .
Figure~\ref{fig:pec_mot} shows the location of W43, together with the
newly determined trace of the Scutum arm and the long Galactic bar. This
suggests that W43 is located close to the meeting point of the Scutum
arm and the near end of the bar.

Three-dimensional motions of maser sources in the Galaxy can be derived
from their measured parallaxes and proper motions, in combination with
their LSR velocities.
Given values for the Galactic parameters $R_0$ (the distance from the
Sun to the Galactic Center) and $\Theta_0$ (the rotation speed of the
Galaxy at the LSR), and assuming a rotation curve, we can estimate the
peculiar motions (i.e., non-circular motions) of the masers by
subtracting the effects of Galactic rotation and peculiar motion of the
Sun~\citep{2009ApJ...700..137R}.
We adopt the recently derived Galactic parameters ($R_0$ = 8.34 $\pm$
0.15 kpc, $\Theta_0$ = 246 $\pm$ 5 \kms) and the Solar Motion values
(\U\ =  10.5 $\pm$ 1.7 \kms, \V\ = 11.3 $\pm$ 2.0 \kms, and \W\ =  8.8
$\pm$ 1.0 \kms) by \citealt{2013Reid} (``Model B1''), to estimate the
peculiar motions of sources in W43.  Since the Galactocentric distance
of W43 is about 4.5 kpc and a flat rotation curve might be not valid
inside of 5 kpc from the Galactic Center, we adopt a non-flat rotation
curve $\Theta (R) = 205 + 10\times(R-2)$ \kms\ for the inner Galaxy
($2\le R \le 5$ kpc), which is adapted from Figure 1 in
\citet{2013RAA....13..849X}.

The peculiar motions for the four maser sources in W43 listed in
Table~\ref{tab:pecmot}, show that motions out of the Galactic plane
(toward the north Galactic pole, \Ws) are small, as expected for a
massive star-forming region whose Galactic orbit should be mainly in the
Galactic plane.  By averaging the results of the four maser sources, W43
shows a small peculiar motion in the direction of Galactic rotation
(\Vs) but a large motion toward the Galactic Center (\Us).  As suggested
by \citet{1979ApJ...233...67R},  this non-circular motion toward the
Galactic Center and along the orientation of the Galactic bar, is likely
induced by the gravitational potential of the central bar.  Furthermore,
\citet{1979ApJ...233...67R} also suggested that in regions of
convergence where the spiral arm bends from the bar, shocks focus gas in
the inner parts outward and gas in the outer part inward.

\section{SUMMARY}

We have measured the parallaxes and proper motions of three 12 GHz
methanol maser sources and one 22 GHz water maser source in the massive
star-forming complex W43.  The distances to these sources are accurate
to 11\% to 18\%: $6.21_{-0.69}^{+0.88}$ kpc for \GIXa,
$5.26_{-0.50}^{+0.62}$ kpc for \GIXb, $4.27_{-0.61}^{+0.85}$ kpc for
\GIa, $5.46_{-0.81}^{+1.16}$ kpc for \GIb.  All of these sources are
consistent with belonging to a single giant molecular cloud, and we
derived the distance of W43 to be $5.49_{-0.34}^{+0.39}$ kpc.  This
locates W43 at the meeting point of the Scutum arm and the Galactic bar.
The peculiar motion of W43 is mainly pointing toward the Galactic Center
along the orientation of the Galactic bar, which is likely induced by
the gravitational attraction of the Galactic bar.

\acknowledgements
{

The work was supported by the National Science Foundation of China
(under grants 10921063, 11073046, 11073054 and 11133008) and the Key
Laboratory for Radio Astronomy, Chinese Academy of Sciences.  This work
was partially funded by the ERC Advanced Investigator Grant GLOSTAR
(247078).  We are grateful to Dr. James Urquhart for providing the
ATLASGAL FITS files.

}

{\it Facilities:} \facility{VLBA}


\appendix
\twocolumn

\section{APPENDIX}
\label{sec:app}

\begin{figure}[H]
  \centering
  \includegraphics[angle=-0,scale=0.70]{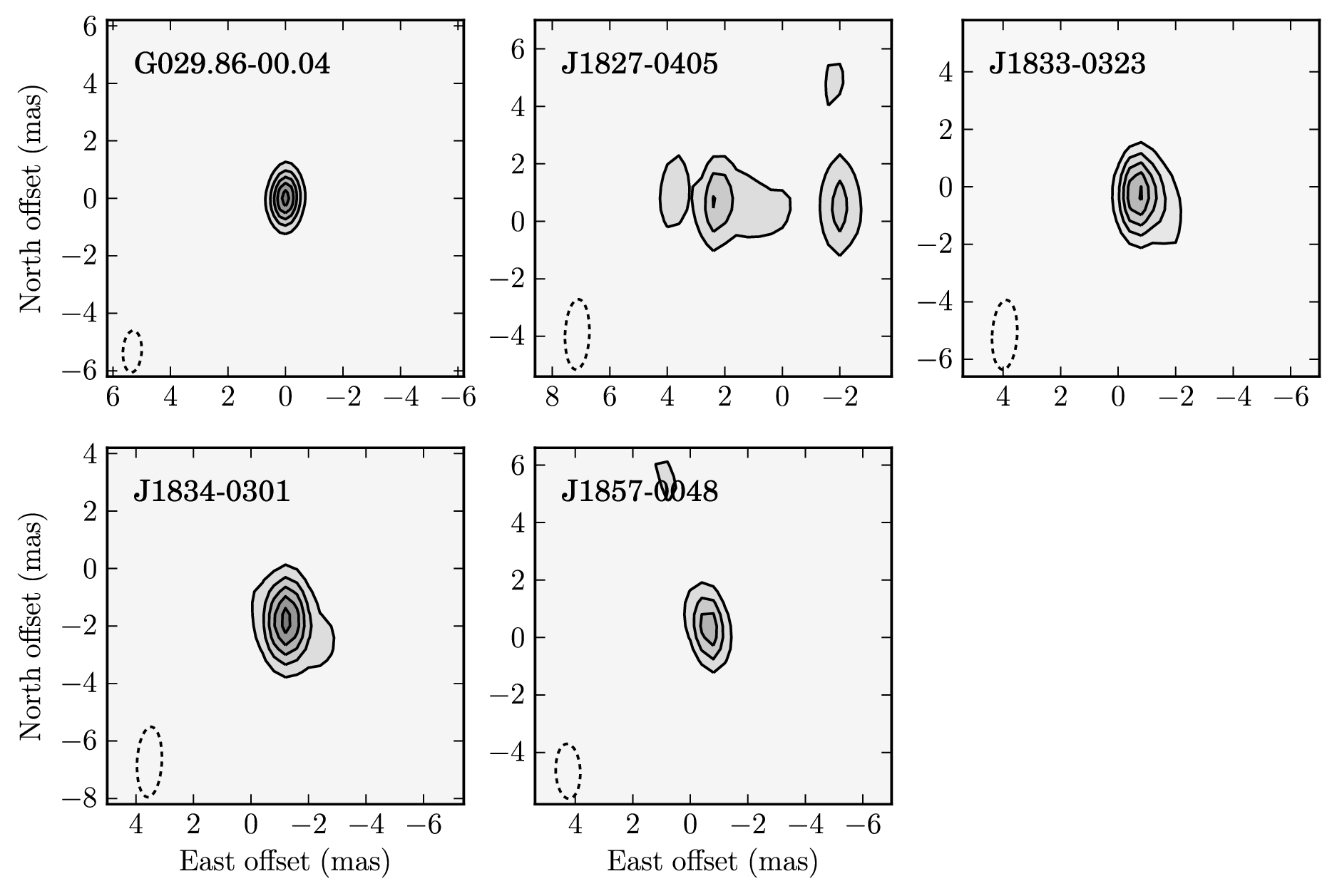}
  \caption{
Images of the 12 GHz \meth\ maser reference spot at \VLSR\ of 100.0
\kms\ and the extragalactic radio sources used for the parallax
measurements of \GIXa\ at the first epoch of the VLBA program BR145Q
(2011 March 20).  Source names are in the upper left corner and the
restoring beam (dotted ellipse) is shown in the lower left corner of each
panel.  Contour levels for the \GIXa\ maser emission are spaced linearly
by 0.5 \jybeam, and for the emission of background sources (\JVII, \JIII, \Jfour\ and
\Jseven), the contours are spaced linearly by 0.010, 0.010, 0.015, and 0.010 \jybeam,
respectively.
  \label{fig:g029a_mq}}
\end{figure}

\begin{figure}[H]
  \centering
  \includegraphics[angle=-0,scale=0.70]{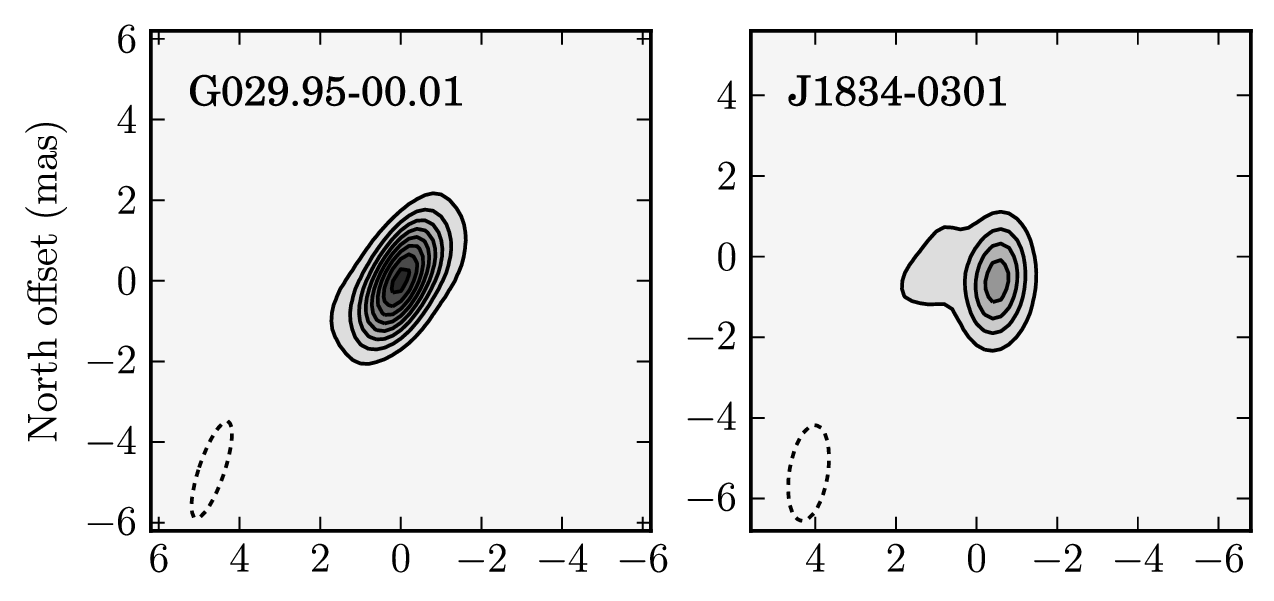}
  \caption{
Images of the 12 GHz \meth\ maser reference spot at \VLSR\ of 96.6 \kms\
and the extragalactic radio sources used for the parallax measurements
of \GIXb\ at the first epoch of the VLBA program BR145W (2011 October
13).  Source names are in the upper left corner and the restoring beam
(dotted ellipse) is shown in the lower left corner of each panel.  Contour
levels for the \GIXb\ maser emission and the \Jfour\ emission are spaced linearly
by 1.0 and 0.020 \jybeam, respectively.
  \label{fig:g029b_mq}}
\end{figure}

\begin{figure}[H]
  \centering
  \includegraphics[angle=-0,scale=0.70]{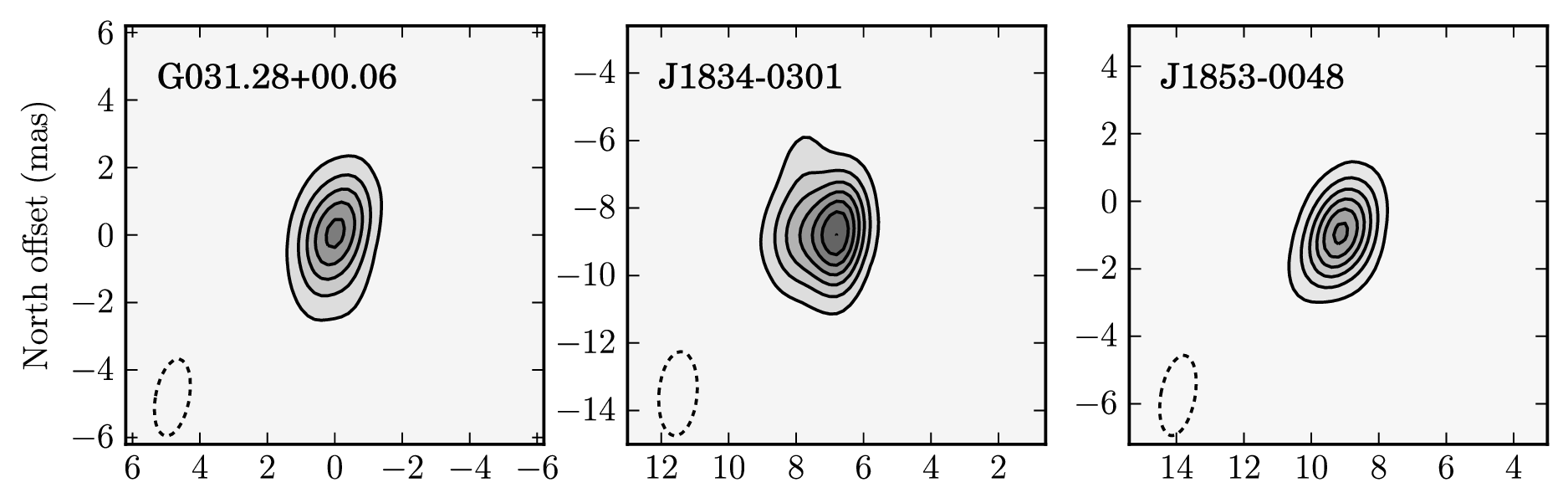}
  \caption{
Images of the 12 GHz \meth\ maser reference spot at \VLSR\ of 99.4 \kms\
and the extragalactic radio sources used for the parallax measurements
of \GIa\ at the last epoch of the VLBA program BR145W (2011 October 13).
Source names are in the upper left corner and the restoring beam (dotted
ellipse) is shown in the lower left corner of each panel.  Contour levels for
the \GIa\ maser emission are spaced linearly by 0.5 \jybeam, and for the
emission of the
background sources (\Jfour\ and \Jthree), the contours are spaced
linearly by 0.006
and 0.003 \jybeam, respectively.
  \label{fig:g031a_mq}}
\end{figure}

\begin{figure}[H]
  \centering
  \includegraphics[angle=-0,scale=0.70]{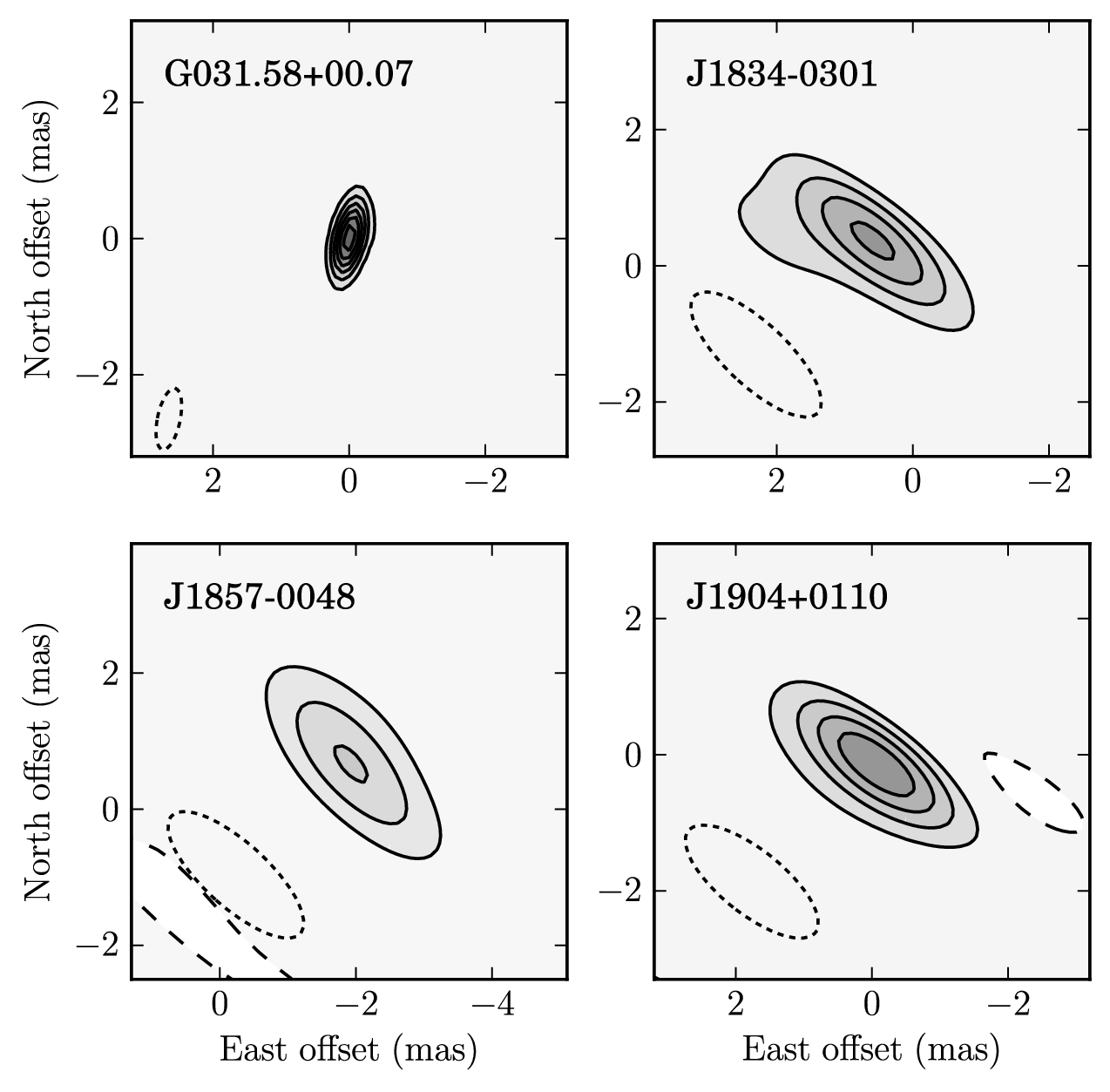}
  \caption{
Images of the 22 GHz \hho\ maser reference spot at \VLSR\ of 110.4 \kms\
and the extragalactic radio sources used for the parallax measurements
of \GIb\ at the first epoch of the VLBA program BR145O (2011 April 03).
Source names are in the upper left corner and the restoring beam (dotted
ellipse) is shown in the lower left corner of each panel.  Contour levels for
\GIb\ maser emission are spaced linearly by 1.5 \jybeam, and for the
emission of the
background sources (\Jfour, \Jseven, and \JIV), the contours are spaced linearly at
0.010, 0.003, and 0.006 \jybeam, respectively.
  \label{fig:g031b_mq}}
\end{figure}

\begin{figure}[H]
  \centering
  \includegraphics[angle=-0,scale=0.70]{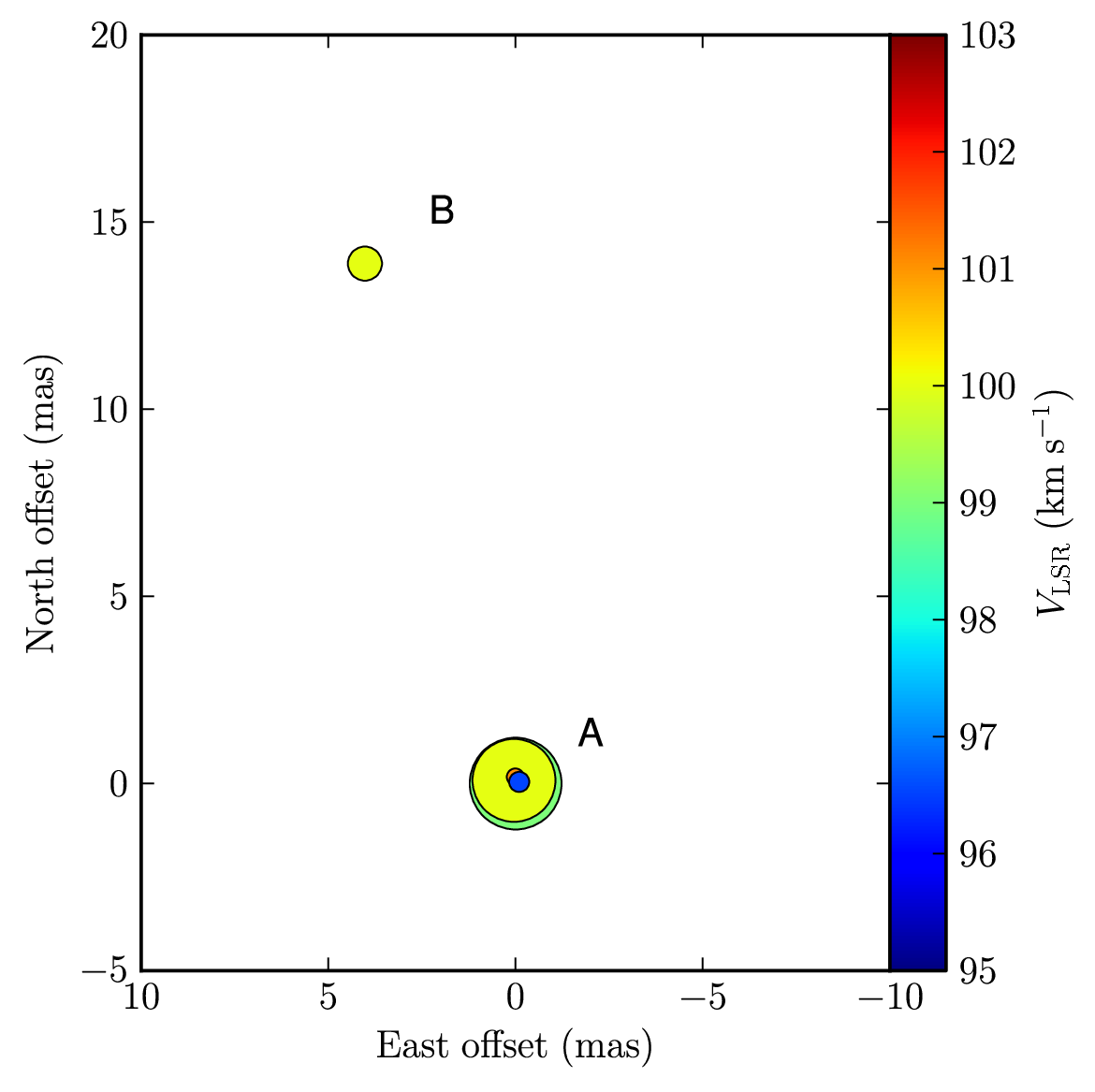}
  \caption{
Spatial distribution of the 12 GHz \meth\ maser spots
toward \GIXa\ at the first epoch of the VLBA
program BR145Q (2011 March 20).  Each maser spot is represented by a
filled circle whose area is proportional to the logarithm of the flux
density.  Each region including maser spots used for parallax fitting is
labeled with a letter.  The reference maser spot is located in region A.
Colors denote maser \VLSR\ with the color-velocity conversion code
given in the wedge on the right of the panel.
(A color
version of this figure is available in the online journal.)
  \label{fig:g029a_spotmap}}
\end{figure}

\begin{figure}[H]
  \centering
  \includegraphics[angle=-0,scale=0.65]{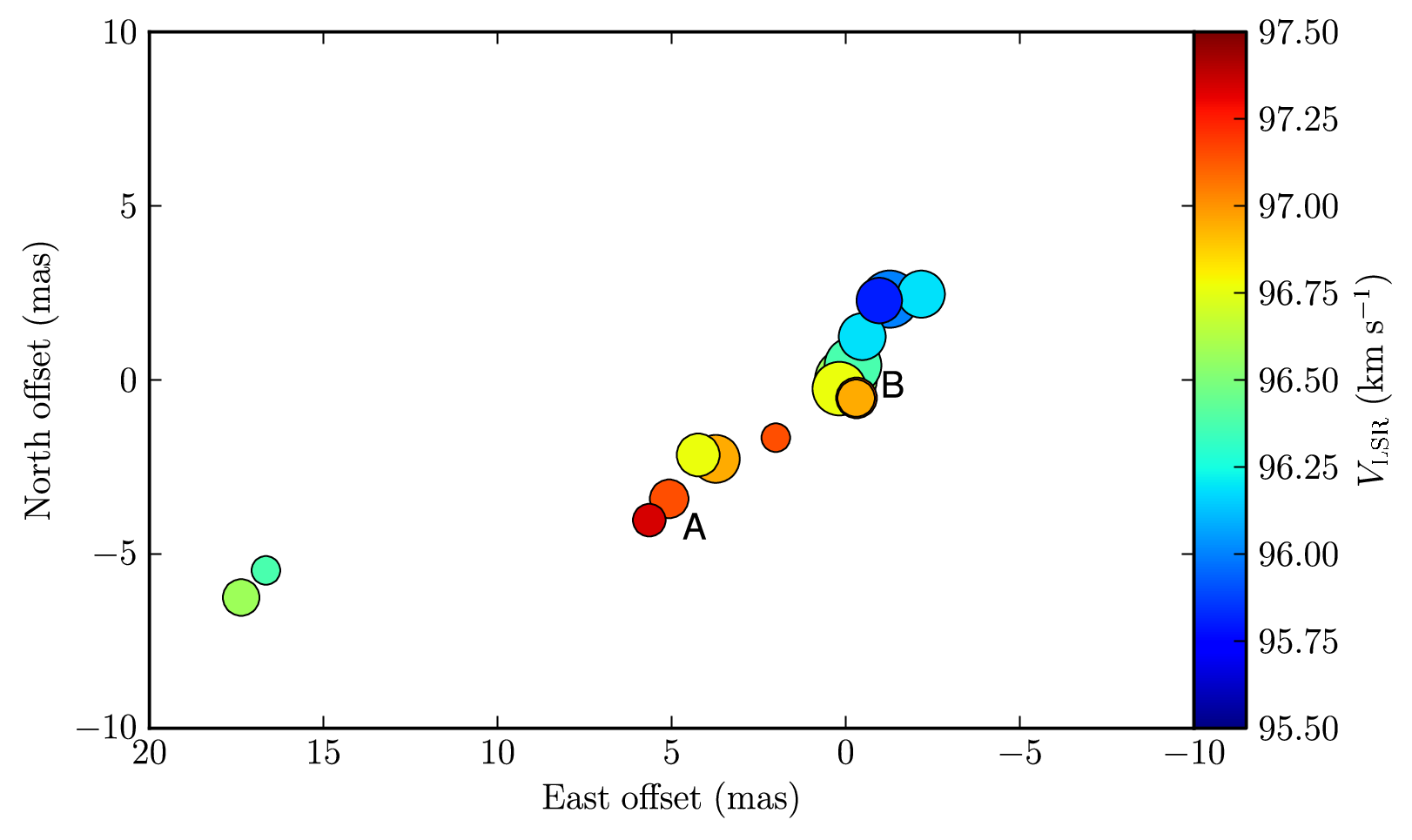}
  \caption{
Spatial distribution of the 12 GHz \meth\ maser spots toward \GIXb\ from
VLBA observations at the first epoch of the VLBA program BR145W (2011
October 13).  Each maser spot is represented by a filled circle whose
area is proportional to the logarithm of the flux density.  Each region
including maser spots used for parallax fitting is labeled with a letter.
The reference maser spot is located in region B.
Colors denote maser \VLSR\ with the color-velocity conversion code
given in the wedge on the right of the panel.
 (A color version of this figure
is available in the online journal.)
  \label{fig:g029b_spotmap}}
\end{figure}

\begin{figure}[H]
  \centering
  \includegraphics[angle=-0,scale=0.70]{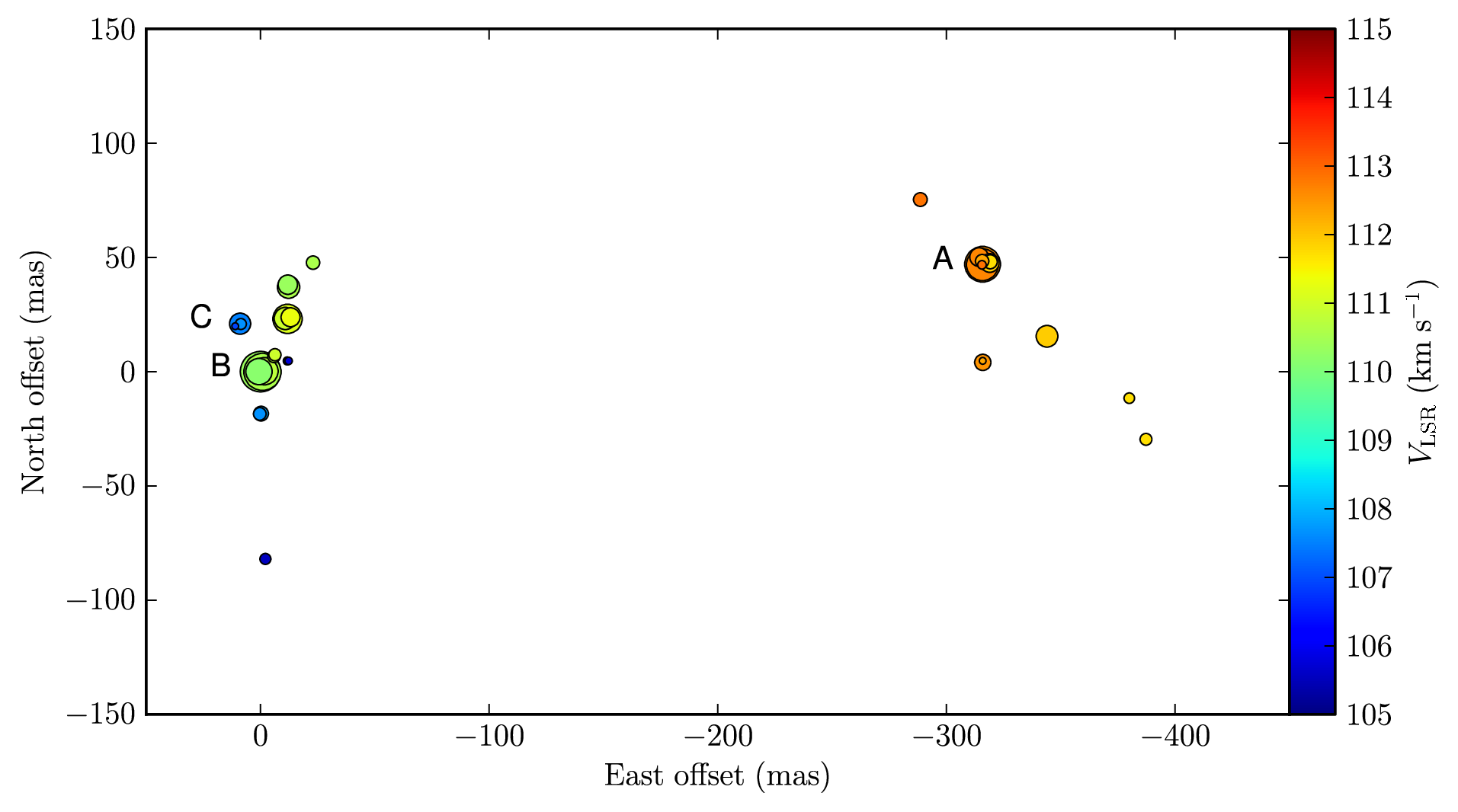}
  \caption{
Spatial distribution of the 12 GHz \meth\ maser spots with brightness
toward \GIa\ at the first epoch of the VLBA
program BR145W (2011 October 13).  Each maser spot is represented by a
filled circle whose area is proportional to the logarithm of the flux
density.  Each region including maser features used for parallax fitting is
labeled with a letter.  The reference maser spot is located in region B.
Colors denote maser \VLSR\ with the color-velocity conversion code
given in the wedge on the right of the panel.
(A color
version of this figure is available in the online journal.)
  \label{fig:g031a_spotmap}}
\end{figure}

\begin{figure}[H]
  \centering
  \includegraphics[angle=-0,scale=0.65]{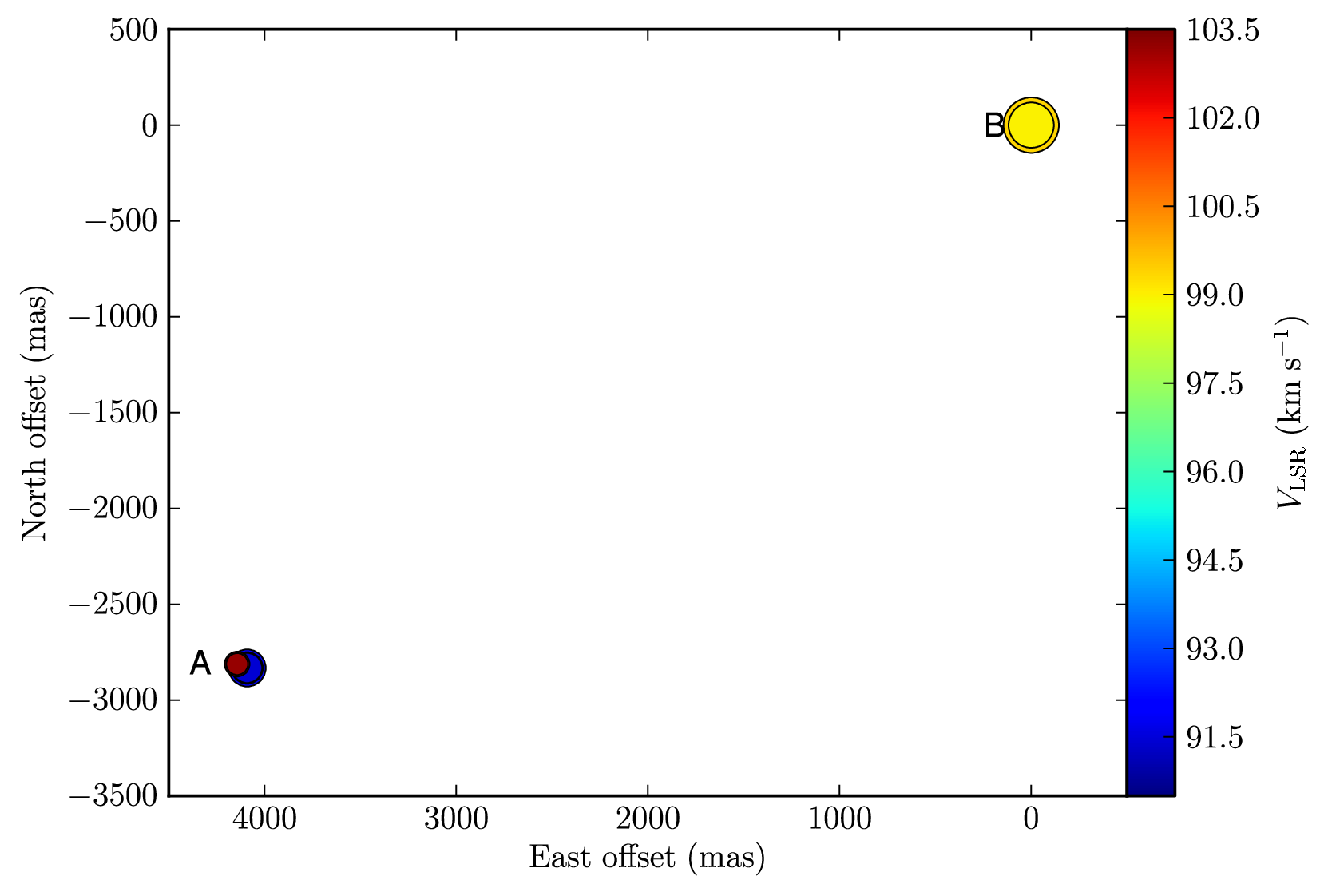}
  \caption{
Spatial distribution of the 22 GHz \hho\ maser spots
toward \GIb\ at the first epoch of the VLBA
program BR145O (2011 April 03).  Each maser spot is represented by a
filled circle whose area is proportional to the logarithm of the flux
density.  Each region including maser features used for parallax fitting is
labeled with a letter.  The reference maser spot is located in region B.
Colors denote maser \VLSR\ with the color-velocity conversion code
given in the wedge on the right of the panel.
(A color
version of this figure is available in the online journal.)
  \label{fig:g031b_spotmap}}
\end{figure}

In Section~\ref{sec:rst}, we outlined the procedures for the parallax
and proper motion measurements.  Here, we describe the details of the
analysis for individual sources.  In the online material, we present the
images of the reference maser spot and the background sources in
Figures~\ref{fig:g029a_mq} -- \ref{fig:g031b_mq}, and the distribution
of the maser spots in Figures~\ref{fig:g029a_spotmap} --
\ref{fig:g031b_spotmap}.  The individual and combined parallax fits are
listed in Tables~\ref{tab:g029a_para_pm} -- \ref{tab:g031b_para_pm}.

\subsection{\GIXa}
\label{ssec:g29a}

For the parallax measurement of \GIXa, we phase-referenced to the 12 GHz
\meth\ maser spot at \VLSR\ of 100.0 \kms.  All four background sources
(\Jfour, \JIII, \Jseven, and \JVII) were detected at all four epochs of
VLBA program BR145Q.    However, since \JVII\ is well separated
(4.7\deg) from the maser source, residual atmospheric effects are
expected to significantly degrade astrometric accuracy.  Hence, we do
not use the data from this background source.

The combined parallax estimate of \GIXa\ is $0.161~\pm~0.020$ mas,
corresponding to a distance of $6.21^{+0.88}_{-0.69}$~kpc.  The absolute
proper motion of \GIXa\ is estimated to be \mux\ = --2.31 $\pm$ 0.01
\masy\ and \muy\ = --5.28 $\pm$ 0.03 \masy, where $\mu_x =
\mu_{\alpha}\cos\delta$ is in the eastward direction and $\mu_y =
\mu_{\delta}$ is in the northward direction.

\subsection{\GIXb}
\label{ssec:g29b}

For the parallax measurement of \GIXb, we used the 12 GHz \meth\ maser
spot at \VLSR\ of 96.6 \kms\ as the interferometer phase reference.  The
background source \Jfour\ was detected at all four epochs of VLBA
program BR145W.

The combined parallax estimate of \GIXa\ is $0.190~\pm~0.020$ mas,
corresponding to a distance of $5.26^{+0.62}_{-0.50}$~kpc.  The absolute
proper motion of \GIXb\ is estimated to be \mux\ = --2.30 $\pm$ 0.03
\masy\ and \muy\ = --5.34 $\pm$ 0.03 \masy.

\subsection{\GIa}
\label{ssec:g31a}

For the parallax measurement of \GIa, we used the 12 GHz \meth\ maser
spot at \VLSR\ of 99.4 \kms\ as the interferometer phase reference.  The
two background sources (\Jfour\ and \Jthree) were detected at all four
epochs of VLBA program BR145W.  The extragalactic source with the better
known position, \Jfour, was relatively far (3\decdeg8) from \GIa; as
such we used it to determine the absolute position of the maser
reference spot, but not for the parallax measurement for the same reason
as described in Section~\ref{ssec:g29a}.  So, the parallax for \GIa\
used only \Jthree, which is separated by only 1\decdeg5 from the maser.

The combined parallax estimate of \GIa\ is $0.234~\pm~0.039$ mas,
corresponding to a distance of $4.27^{+0.87}_{-0.64}$~kpc.   The
absolute proper motion of \GIa\ is estimated to be \mux\ = --2.09 $\pm$
0.05 \masy\ and \muy\ =--4.37 $\pm$ 0.14 \masy.

\subsection{\GIb}
\label{ssec:g31b}

For the parallax measurement of \GIb, we used the 22 GHz \hho\ maser
spot at \VLSR\ of 96.6 \kms\ as the interferometer phase reference.  The
three background sources (\Jfour, \Jseven\ and \JIV) were detected at
all six epochs of VLBA program BR145O.  The two background sources
\Jfour\ and \JIV\ are relatively far away ($>$ 4\deg) from \GIb, and
were not used for parallax estimation.  We used \Jseven, which is
separated by 2\decdeg3 on the sky from the maser source, to estimate the
parallax.

The combined parallax estimate of \GIb\ is $0.183~\pm~0.032$ mas,
corresponding to a distance of $5.46^{+1.16}_{-0.81}$~kpc.  The absolute
proper motion of \GIb\ is estimated to be \mux\ = --2.17 $\pm$ 0.04
\masy\ and \muy\ = --4.50 $\pm$ 0.10 \masy.



\clearpage
\onecolumn


\begin{deluxetable}{cccccl}
  \tablecolumns{5}
  \tablewidth{0pc}
  \tablecaption{VLBA Observations}
  \tablehead{

  \colhead{Program} & \colhead{Source} & \colhead{Maser}   & \colhead{Frequency} & \colhead{\VLSR\ resolution} & \colhead{Epochs} \\
   \colhead{Code}   &  \colhead{Name}  & \colhead{Species} & \colhead{(GHz)}     & \colhead{(\kms)}            & \colhead{(20yymmdd)}

  }
  \startdata

BR145Q & \GIXa & \meth & 12.2 & 0.77 & 110320, 110915, 110926, 120324 \\
       &       &       &      &      &                                \\
BR145W & \GIXb & \meth & 12.2 & 0.19 & 111013, 120318, 120410, 120924 \\
BR145W & \GIa  & \meth & 12.2 & 0.19 &                                \\
       &       &       &      &      &                                \\
BR145O & \GIb  & \hho  & 22.2 & 0.42 & 110403, 110628, 110905, 111020, 111206, 120404 \\

\enddata
\label{tab:obs}
\end{deluxetable}

\begin{deluxetable}{ccccrrrc}
  \tablecolumns{8}
  \tablewidth{0pc}
  \tablecaption{Source characteristics}
  \tablehead{

  \colhead{Source} & \colhead{R.A. (J2000)} & \colhead{Dec. (J2000)}                & \colhead{$\theta_{sep}$} & \colhead{P.A.}    & \colhead{$S$}            & \colhead{\VLSR}  & \colhead{Beam}              \\
  \colhead{     }  & \colhead{(h~~~m~~~s)}  & \colhead{(\degr~~~\arcmin~~~\arcsec)} & \colhead{(\degr)}        & \colhead{(\degr)} & \colhead{(Jy~bm$^{-1}$)} & \colhead{(\kms)} & \colhead{(mas~~mas~~\degr)} \\
   \colhead{(1)}   & \colhead{(2)}          & \colhead{(3)}                         & \colhead{(4)}            & \colhead{(5)}     & \colhead{(6)}            & \colhead{(7)}    & \colhead{(8)}

  }
  \startdata

      \GIXa & 18 45 59.5690 & $-$02 45 06.408 & ... &    ... &  2.7\phn &    100.0 & 1.4 $\times$ 0.7 @ \phn$-$3  \\
     \Jfour & 18 34 14.0746 & $-$03 01 19.629 & 2.9 &  $-$95 &     0.09 &      ... & 2.5 $\times$ 0.9 @ \phn$-$3  \\
      \JIII & 18 33 23.9044 & $-$03 23 31.432 & 3.2 & $-$102 &     0.05 &      ... & 2.4 $\times$ 0.9 @ \phn$-$3  \\
    \Jseven & 18 57 51.3588 & $-$00 48 21.932 & 3.5 &    +57 &     0.04 &      ... & 1.9 $\times$ 0.8 @ \phn  +2  \\
      \JVII & 18 27 45.0404 & $-$04 05 44.580 & 4.7 & $-$106 &     0.02 &      ... & 2.4 $\times$ 0.8 @ \phn$-$2  \\
            &               &                 &     &        &          &          &                              \\
      \GIXb & 18 46 03.7402 & $-$02 39 22.328 & ... &    ... &  8.5\phn &     96.4 & 2.5 $\times$ 0.6 @    $-$18  \\
     \Jfour & 18 34 14.0746 & $-$03 01 19.627 & 3.0 &  $-$97 &     0.09 &      ... & 2.4 $\times$ 0.9 @ \phn$-$9  \\
            &               &                 &     &        &          &          &                              \\
       \GIa & 18 48 12.3900 & $-$01 26 30.695 & ... &    ... &  2.6\phn &    110.4 & 2.3 $\times$ 1.1 @    $-$11  \\
    \Jthree & 18 53 41.9892 & $-$00 48 54.330 & 1.5 &    +65 &     0.02 &      ... & 2.4 $\times$ 1.0 @    $-$10  \\
     \Jfour & 18 34 14.0746 & $-$03 01 19.627 & 3.8 & $-$114 &     0.04 &      ... & 2.5 $\times$ 1.1 @ \phn$-$5  \\
            &               &                 &     &        &          &          &                              \\
       \GIb & 18 48 41.6740 & $-$01 09 59.774 & ... &    ... & 10.1\phn &     99.4 & 0.9 $\times$ 0.3 @    $-$13  \\
    \Jseven & 18 57 51.3589 & $-$00 48 21.938 & 2.3 &    +81 &     0.01 &      ... & 2.6 $\times$ 0.9 @      +48  \\
     \Jfour & 18 34 14.0746 & $-$03 01 19.627 & 4.1 & $-$117 &     0.04 &      ... & 2.5 $\times$ 0.9 @      +47  \\
       \JIV & 19 04 26.3977 &   +01 10 36.698 & 4.6 &    +59 &     0.03 &      ... & 2.4 $\times$ 0.9 @      +51  \\

\enddata
\tablecomments{
Column 1 gives the names of the maser sources and the corresponding
background sources. Columns 2 to 3 list the absolute positions of the
reference maser spot and the background sources.  Columns 4 to 5 give
the separations ($\theta_{sep})$ and position angles (P.A. east of
north) between maser and background sources.  Columns 6 to 7 give the
source brightnesses ($S$) and \VLSR\ of the reference maser spot.
Column 8 gives the full width at half maximum (FWHM) size and P.A. of
the Gaussian restoring beam.
Calibrator \Jfour\ is from the third VLBA Calibrator Survey
(VCS)~\citep{2005AJ....129.1163P}; \Jthree, \JIII, and \JVII\ are from
the BeSSeL calibrator survey \citep{2011ApJS..194...25I}; \Jseven\ is
from a targeted Very Large Array calibrator search
\citep{2006ApJS..166..526X}; \JIV\ is from the
ICRF2~\citep{2009ITN....35....1M}.
}
\label{tab:src}
\end{deluxetable}

\begin{deluxetable}{cccllr}
\tablecolumns{6}
\tablewidth{0pc}
\tabletypesize{\footnotesize} 
\tablecaption{Parallaxes and proper motions of maser sources in W43}
\tablehead{

\colhead{Source}  & \colhead{Parallax} & \colhead{Distance} & \colhead{\mux}    & \colhead{\muy}    & \colhead{\VLSR} \\
\colhead{name}    & \colhead{(mas)}    & \colhead{(kpc)}    & \colhead{(\masy)} & \colhead{(\masy)} & \colhead{(\kms)} \\
\colhead{(1)}    & \colhead{(2)}     & \colhead{(3)}      & \colhead{(4)}      & \colhead{(5)}     & \colhead{(6)}

}

\startdata

\GIXa & 0.161 $\pm$ 0.020 & $6.21_{-0.69}^{+0.88}$ & $-$2.30 $\pm$ 0.01 (0.17) & $-$5.32 $\pm$ 0.02 (0.17) &    100 $\pm$ 3 \\
\GIXb & 0.190 $\pm$ 0.020 & $5.26_{-0.50}^{+0.62}$ & $-$2.32 $\pm$ 0.02 (0.20) & $-$5.38 $\pm$ 0.02 (0.20) & \phn98 $\pm$ 3 \\
 \GIa & 0.234 $\pm$ 0.039 & $4.27_{-0.61}^{+0.85}$ & $-$2.09 $\pm$ 0.03 (0.25) & $-$4.38 $\pm$ 0.13 (0.28) &    109 $\pm$ 3 \\
 \GIb & 0.183 $\pm$ 0.032 & $5.46_{-0.81}^{+1.16}$ & $-$1.97 $\pm$ 0.01 (0.19) & $-$4.43 $\pm$ 0.10 (0.21) & \phn96 $\pm$ 5 \\
&&& && \\
W43   & 0.182 $\pm$ 0.012 & $5.49_{-0.34}^{+0.39}$ & $-$2.18 $\pm$ 0.10       & $-$5.32 $\pm$ 0.10       &    102 $\pm$ 2 \\

\enddata

\tablecomments{
Columns 2 and 3 lists the parallax value and the derived source
distance.  Column 4 to 6 list absolute proper motions in the eastward
and northward direction and \VLSR, respectively.  For proper motion we
report two errors. The first is is the measured uncertainty from the
parallax fits. The second in parenthesis is derived by adding in
quadrature to the first an error floor of 5 \kms, to account for the
uncertainty in our estimate of the reference system of the star.  In the
last line, we list the variance-weighted parallax, proper motion, and
\VLSR\ of W43, derived from the four sources listed above.
}

\label{tab:ppm}
\end{deluxetable}

\begin{deluxetable}{cccc}
\tablecolumns{4}
\tablewidth{0pc}
\tabletypesize{\footnotesize} 
\tablecaption{Peculiar motions of maser sources in W43}
\tablehead{

\colhead{Source} & \colhead{\Us}    & \colhead{\Vs}    & \colhead{\Ws} \\
\colhead{name}   & \colhead{(\kms)} & \colhead{(\kms)} & \colhead{(\kms)} \\
\colhead{(1)}    & \colhead{(2)}    & \colhead{(3)}    & \colhead{(4)}

}

\startdata

\GIXa & \phn6.5 $\pm$  22.2 &  --2.1  $\pm$  \phn6.0 &   --2.5 $\pm$ 5.5\\
\GIXb &    28.4 $\pm$  13.7 &  --2.3  $\pm$  \phn6.1 &   --1.0 $\pm$ 5.3\\
 \GIa &    32.2 $\pm$  22.6 &   23.1  $\pm$  \phn7.5 & \phn6.1 $\pm$ 5.5\\
 \GIb &   --4.3 $\pm$  29.6 &\phn8.1  $\pm$     11.6 & \phn2.0 $\pm$ 5.6\\
& & & \\
  W43 &    20.5 $\pm$  10.7 &\phn3.8  $\pm$  \phn5.1 &   --4.0 $\pm$ 2.9\\

\enddata

\tablecomments{
Columns 2 to 4 list peculiar motion components, where \Us, \Vs, \Ws\ are
directed toward the Galactic Center, in the direction of Galactic
rotation and toward the North Galactic Pole (NGP), respectively. The
peculiar motions were estimated using the Galactic parameters ($R_0$ =
8.34 $\pm$ 0.15 kpc, $\Theta_0$ = 246 $\pm$ 5 \kms) and the Solar Motion
values (\U\ =  10.5 $\pm$ 1.7 \kms, \V\ = 11.3 $\pm$ 2.0 \kms, and \W\ =
8.8 $\pm$ 1.0 \kms) from by M.  J.  Reid et al. (2013, in preparation,
``Model B1''), assuming a rotation curve ($\Theta (R) = 205 +
10\times(R-2)$ \kms) for the inner Galaxy ($2\le R \le 5$ kpc) adapted
from \citet{2013RAA....13..849X}.
}

\label{tab:pecmot}
\end{deluxetable}

\clearpage

\section{ONLINE MATERIAL}

\setlength{\tabcolsep}{2pt}
\begin{deluxetable}{lccrrrrrrcc}
\tabletypesize{\footnotesize} 
\tablecaption{Parallax and proper motion fits for \GIXa~\label{tab:g029a_para_pm}}
\tablewidth{0pt}
\tablehead{ 
\colhead{Background} & \colhead{Region} & \colhead{\VLSR}  & \colhead{Parallax} & \colhead{\mux}    & \colhead{\muy}    & \colhead{\dx}   & \colhead{\dy}   & \colhead{$\chi^2_{\nu}$} & \colhead{$\sigma_x$} & \colhead{$\sigma_y$} \\
\colhead{source}     & \colhead{}       & \colhead{(\kms)} &
\colhead{(mas)}    & \colhead{(\masy)} & \colhead{(\masy)} &
\colhead{(mas)} & \colhead{(mas)} & \colhead{}               &
\colhead{(mas)}             & \colhead{(mas)}
}
\startdata

   \JIII  &  A  & 102.31  & 0.119 $\pm$ 0.005  & --2.22  $\pm$ 0.01  & --5.33  $\pm$ 0.05  & --0.53  $\pm$ 0.00  & --2.61  $\pm$ 0.02  & 0.933  & 0.000  &  0.038\\
          &  A  & 100.77  & 0.117 $\pm$ 0.004  & --2.23  $\pm$ 0.01  & --5.34  $\pm$ 0.07  & --0.53  $\pm$ 0.00  & --2.43  $\pm$ 0.02  & 0.561  & 0.000  &  0.066\\
          &  B  & 100.77  & 0.103 $\pm$ 0.005  & --2.27  $\pm$ 0.01  & --5.32  $\pm$ 0.14  &   3.44  $\pm$ 0.01  &  11.35  $\pm$ 0.05  & 0.998  & 0.002  &  0.099\\
          &  A  & 100.00  & 0.119 $\pm$ 0.006  & --2.22  $\pm$ 0.02  & --5.33  $\pm$ 0.05  & --0.56  $\pm$ 0.01  & --2.51  $\pm$ 0.02  & 0.997  & 0.005  &  0.035\\
          &  A  &  98.46  & 0.123 $\pm$ 0.011  & --2.20  $\pm$ 0.03  & --5.22  $\pm$ 0.15  & --0.56  $\pm$ 0.01  & --2.44  $\pm$ 0.05  & 0.997  & 0.019  &  0.107\\
          &     &         &                    &                     &                     &                    &                    &        &        &       \\
  \Jfour  &  A  & 102.31  & 0.194 $\pm$ 0.025  & --2.23  $\pm$ 0.07  & --5.32  $\pm$ 0.08  & --0.04  $\pm$ 0.02  & --1.06  $\pm$ 0.03  & 1.000  & 0.047  &  0.058\\
          &  A  & 100.77  & 0.192 $\pm$ 0.018  & --2.24  $\pm$ 0.05  & --5.32  $\pm$ 0.05  & --0.05  $\pm$ 0.02  & --0.88  $\pm$ 0.02  & 1.000  & 0.034  &  0.033\\
          &  B  & 100.77  & 0.178 $\pm$ 0.015  & --2.27  $\pm$ 0.04  & --5.31  $\pm$ 0.11  &   3.93  $\pm$ 0.01  &  12.90  $\pm$ 0.04  & 0.998  & 0.027  &  0.076\\
          &  A  & 100.00  & 0.194 $\pm$ 0.026  & --2.23  $\pm$ 0.07  & --5.32  $\pm$ 0.09  & --0.08  $\pm$ 0.03  & --0.96  $\pm$ 0.03  & 1.000  & 0.049  &  0.061\\
          &  A  &  98.46  & 0.198 $\pm$ 0.031  & --2.20  $\pm$ 0.09  & --5.21  $\pm$ 0.28  & --0.07  $\pm$ 0.03  & --0.89  $\pm$ 0.10  & 0.998  & 0.060  &  0.201\\
          &     &         &                    &                     &                     &                    &                    &        &        &       \\
 \Jseven  &  A  & 102.31  & 0.179 $\pm$ 0.007  & --2.46  $\pm$ 0.02  & --5.26  $\pm$ 0.43  & --0.66  $\pm$ 0.01  & --3.00  $\pm$ 0.15  & 1.000  & 0.008  &  0.305\\
          &  A  & 100.77  & 0.177 $\pm$ 0.013  & --2.47  $\pm$ 0.04  & --5.27  $\pm$ 0.39  & --0.67  $\pm$ 0.01  & --2.82  $\pm$ 0.14  & 1.000  & 0.024  &  0.282\\
          &  B  & 100.77  & 0.163 $\pm$ 0.017  & --2.50  $\pm$ 0.05  & --5.25  $\pm$ 0.42  &   3.31  $\pm$ 0.02  &  10.96  $\pm$ 0.15  & 1.000  & 0.031  &  0.301\\
          &  A  & 100.00  & 0.179 $\pm$ 0.006  & --2.46  $\pm$ 0.02  & --5.26  $\pm$ 0.43  & --0.69  $\pm$ 0.01  & --2.89  $\pm$ 0.15  & 1.000  & 0.005  &  0.307\\
          &  A  &  98.46  & 0.183 $\pm$ 0.004  & --2.43  $\pm$ 0.01  & --5.15  $\pm$ 0.43  & --0.69  $\pm$ 0.00  & --2.83  $\pm$ 0.15  & 0.502  & 0.000  &  0.434\\
          &     &         &                    &                     &                     &                    &                    &        &        &       \\
 Combined & all &         & 0.161 $\pm$ 0.020  &                     &                     &                    &                    & 1.000  & 0.066  & 0.155 \\
          &     &         &                    &                     &                     &                    &                    &        &        &       \\
 Averaged & all &         &                    & --2.30 $\pm$ 0.01   &  --5.32 $\pm$ 0.02  &                    &                    &        &        &       \\

\enddata

\tablecomments{
Absolute proper motions are defined as $\mux = \mu_{\alpha
\cos{\delta}}$ and $\muy = \mu_{\delta}$.  $\chi^2_{\nu}$ is the reduced
$\chi^2$ of post-fit residuals,  $\sigma_x$ and $\sigma_y$ are error
floor in $x$ and $y$, respectively. 
}
\end{deluxetable}

\setlength{\tabcolsep}{2pt}
\begin{deluxetable}{lccrrrrrrcc}
 \tabletypesize{\footnotesize} 
\tablecaption{Parallax and proper motion fits for \GIXb~\label{tab:g029b_para_pm}}
\tablewidth{0pt}
\tablehead{ 
\colhead{Background} & \colhead{Region} & \colhead{\VLSR}  & \colhead{Parallax} & \colhead{\mux}    & \colhead{\muy}    & \colhead{\dx}   & \colhead{\dy}   & \colhead{$\chi^2_{\nu}$} & \colhead{$\sigma_x$} & \colhead{$\sigma_y$} \\
\colhead{source}     & \colhead{}       & \colhead{(\kms)} &
\colhead{(mas)}    & \colhead{(\masy)} & \colhead{(\masy)} &
\colhead{(mas)} & \colhead{(mas)} & \colhead{}               &
\colhead{(mas)}             & \colhead{(mas)}
}
\startdata

    \Jfour   & A   & 97.35 & 0.176 $\pm$ 0.015 & --2.26 $\pm$ 0.04 & --5.39 $\pm$ 0.02 & 5.26 $\pm$ 0.01   & --6.01 $\pm$ 0.01 & 0.998 & 0.030 & 0.012 \\
             & B   & 96.58 & 0.186 $\pm$ 0.025 & --2.31 $\pm$ 0.08 & --5.35 $\pm$ 0.06 & --0.43 $\pm$ 0.03 & --1.93 $\pm$ 0.02 & 0.996 & 0.044 & 0.036 \\
             & B   & 95.81 & 0.210 $\pm$ 0.010 & --2.35 $\pm$ 0.03 & --5.30 $\pm$ 0.07 & --1.38 $\pm$ 0.01 & 0.33 $\pm$ 0.02   & 1.000 & 0.020 & 0.045 \\
             &     &       &                   &                   &                   &                   &                   &       &       & \\
  Combined   & all &       & 0.190 $\pm$ 0.020 &                   &                   &                   &                   & 0.950 & 0.036 & 0.030 \\
             &     &       &                   &                   &                   &                   &                   &       &       & \\
  Averaged   & all &       &                   & --2.32 $\pm$ 0.02 & --5.38 $\pm$ 0.02 &                   &                   &       &       & \\

\enddata

\tablecomments{see~Table~\ref{tab:g029a_para_pm}}

\end{deluxetable}

\setlength{\tabcolsep}{2pt}
\begin{deluxetable}{lccrrrrrrcc}
 \tabletypesize{\footnotesize} 
\tablecaption{Parallax and proper motion fits for \GIa~\label{tab:g031a_para_pm}}
\tablewidth{0pt}
\tablehead{ 
\colhead{Background} & \colhead{Region} & \colhead{\VLSR}  & \colhead{Parallax} & \colhead{\mux}    & \colhead{\muy}    & \colhead{\dx}   & \colhead{\dy}   & \colhead{$\chi^2_{\nu}$} & \colhead{$\sigma_x$} & \colhead{$\sigma_y$} \\
\colhead{source}     & \colhead{}       & \colhead{(\kms)} &
\colhead{(mas)}    & \colhead{(\masy)} & \colhead{(\masy)} &
\colhead{(mas)} & \colhead{(mas)} & \colhead{}               &
\colhead{(mas)}             & \colhead{(mas)}
}
\startdata

  \Jthree &   A & 112.69 & 0.200 $\pm$ 0.044 & --2.19 $\pm$ 0.12 & --4.23 $\pm$ 0.32 & --323.81 $\pm$ 0.04 & 49.97 $\pm$ 0.11 & 0.983 & 0.078 & 0.212\\
          &   A & 112.50 & 0.209 $\pm$ 0.047 & --2.11 $\pm$ 0.13 & --4.31 $\pm$ 0.38 & --323.85 $\pm$ 0.05 & 50.13 $\pm$ 0.13 & 0.980 & 0.085 & 0.255\\
          &   B & 110.38 & 0.257 $\pm$ 0.034 & --1.97 $\pm$ 0.09 & --4.56 $\pm$ 0.35 &   --7.96 $\pm$ 0.03 &  2.99 $\pm$ 0.12 & 0.977 & 0.058 & 0.239\\
          &   C & 107.50 & 0.274 $\pm$ 0.018 & --2.09 $\pm$ 0.03 & --4.39 $\pm$ 0.17 &     0.96 $\pm$ 0.02 & 24.10 $\pm$ 0.06 & 0.520 & 0.000 & 0.153\\
          &     &        &                   &                   &                   &                     &                  &       &       &      \\
 Combined & all &        & 0.234 $\pm$ 0.039 &                   &                   &                     &                  & 0.990 & 0.066 & 0.192\\
          &     &        &                   &                   &                   &                     &                  &       &       &      \\
 Averaged & all &        &                   & --2.09 $\pm$ 0.03 & --4.38 $\pm$ 0.13 &                     &                  &       &       &      \\

\enddata

\tablecomments{see~Table~\ref{tab:g029a_para_pm}}

\end{deluxetable}

\setlength{\tabcolsep}{2pt}
\begin{deluxetable}{lcrrrrrrrcc}
 \tabletypesize{\footnotesize} 
\tablecaption{Parallax and proper motion fits for \GIb~\label{tab:g031b_para_pm}}
\tablewidth{0pt}
\tablehead{ 
\colhead{Background} & \colhead{Region} & \colhead{\VLSR}  & \colhead{Parallax} & \colhead{\mux}    & \colhead{\muy}    & \colhead{\dx}   & \colhead{\dy}   & \colhead{$\chi^2_{\nu}$} & \colhead{$\sigma_x$} & \colhead{$\sigma_y$} \\
\colhead{source}     & \colhead{}       & \colhead{(\kms)} &
\colhead{(mas)}    & \colhead{(\masy)} & \colhead{(\masy)} &
\colhead{(mas)} & \colhead{(mas)} & \colhead{}               &
\colhead{(mas)}             & \colhead{(mas)}
}
\startdata

 \Jseven &   B & 102.79 & 0.176 $\pm$ 0.018 & --2.36 $\pm$ 0.04 & --4.03 $\pm$ 0.16 & 4143.94 $\pm$ 0.01 & --2815.61 $\pm$ 0.05 & 0.998 & 0.021 & 0.121 \\
         &   A &  99.84 & 0.215 $\pm$ 0.009 & --1.91 $\pm$ 0.02 & --4.76 $\pm$ 0.21 &  --1.73 $\pm$ 0.01 &    --9.85 $\pm$ 0.07 & 0.886 & 0.000 & 0.171 \\
         &   A &  99.42 & 0.213 $\pm$ 0.009 & --1.91 $\pm$ 0.02 & --4.74 $\pm$ 0.20 &  --1.73 $\pm$ 0.01 &    --9.85 $\pm$ 0.06 & 0.790 & 0.000 & 0.172 \\
         &   B &  92.26 & 0.153 $\pm$ 0.055 & --2.59 $\pm$ 0.14 & --4.47 $\pm$ 0.23 & 4088.10 $\pm$ 0.04 & --2834.84 $\pm$ 0.07 & 0.998 & 0.105 & 0.171 \\
         &     &        &                   &                   &                   &                    &                      &       &       &       \\
Combined & all &        & 0.183 $\pm$ 0.032 &                   &                   &                    &                      & 1.040 & 0.055 & 0.143 \\
         &     &        &                   &                   &                   &                    &                      &       &       &       \\
Averaged & all &        &                   & --1.97 $\pm$ 0.01 & --4.43 $\pm$ 0.10 &                    &                      &       &       &       \\

\enddata

\tablecomments{see~Table~\ref{tab:g029a_para_pm}.}
\end{deluxetable}

\clearpage


\end{document}